\documentclass[11pt,preprint]{aastex}
\newcommand{\msun}{M_\odot}

\newcommand{\mbh}{\ensuremath{M_{\rm BH}}}

\newcommand{\ledd}{\ensuremath{L_{Edd}}}
\newcommand{\lbol}{\ensuremath{L_{Bol}}}
\newcommand{\llciv}{\ensuremath{L_{1549} }}
\newcommand{\lledd}{\ensuremath{L_{Bol}/L_{Edd}}}

\newcommand{\lllambda}{\ensuremath{\lambda L_{\lambda}}}
\newcommand{\flambda}{\ensuremath{f_{\lambda}}}
\newcommand{\fnu}{\ensuremath{f_{\nu}}}

\newcommand{\nhi}{$N_{\rm HI}$}

\newcommand{\alphal}{{\ensuremath{\alpha_{\lambda}}}}
\newcommand{\alambda}{\ensuremath{\alpha_{\lambda}}}
\newcommand{\anu}{\ensuremath{\alpha_{\nu}}}

\newcommand{\auvoa}{\ensuremath{\alpha_{{UVO}}}}
\newcommand{\auvob}{\ensuremath{\alpha_{{Ored}}}}
\newcommand{\aored}{\ensuremath{\alpha_{{Ored}}}}
\newcommand{\auvo}{\ensuremath{\alpha_{{UVO}}}}

\newcommand{\ax}{{\ensuremath{\alpha_{\rm X}}}}
\newcommand{\alphax}{$\alpha_{\rm X}$}
\newcommand{\aox}{\ensuremath{\alpha_{ox}}}

\newcommand{\dv}{{\ensuremath{\Delta v}}}

\newcommand{\axnu}{\ensuremath{\alpha_{\rm X}}}

\newcommand{\civmshi}{{\civ\ $\Delta v$}}
\newcommand{\hbmshif}{{\hb\ $\Delta v$}}
\newcommand{\hamshif}{{\ha\ $\Delta v$}}
\newcommand{\lyamsh}{{\lya\ $\Delta v$}}
\newcommand{\ciiishi}{{\ciii\ $\Delta v$}}
\newcommand{\mgiishi}{{\mgii\ $\Delta v$}}

\newcommand{\hbasymm}{{\hb\ Asymm}}
\newcommand{\haasymm}{{\ha\ Asymm}}
\newcommand{\civasym}{{\civ\ Asymm}}
\newcommand{\lyaasym}{{\lya\ Asymm}}

\newcommand{\lya}{\ensuremath{{\rm Ly}{\alpha}}}
\newcommand{\lyal}{Ly{\sc $\alpha$}\,$\lambda$1216}

\newcommand{\nv}{N\,{\sc v}}
\newcommand{\nvl}{N\,{\sc v}\,$\lambda$1240}

\newcommand{\niii}{N\,{\sc iii}]}

\newcommand{\niv}{N\,{\sc iv}]}

\newcommand{\ha}{\ensuremath{{\rm H}{\alpha}}}
\newcommand{\hb}{\ensuremath{{\rm H}{\beta}}}

\newcommand{\hi}{H\,{\sc i}}
\newcommand{\siiv}{Si\,{\sc iv}}
\newcommand{\siivl}{Si\,{\sc iv}\,$\lambda$1397}
\newcommand{\oiv}{O\,{\sc iv}]}
\newcommand{\oivl}{O\,{\sc iv}]\,$\lambda$1402}

\newcommand{\siivoivl}{\siiv+\oiv\,$\lambda$1400}
\newcommand{\civ}{C\,{\sc iv}}
\newcommand{\civl}{C\,{\sc iv}\,$\lambda$1549}

\newcommand{\siiii}{Si\,{\sc iii}]}
\newcommand{\siiiil}{Si\,{\sc iii}]\,$\lambda$1892}
\newcommand{\ciii}{{C\,{\sc iii}]}}
\newcommand{\ciiil}{C\,{\sc iii}]\,$\lambda$1909}
\newcommand{\aliii}{Al\,{\sc iii}}

\newcommand{\hei}{He\,{\sc i}}
\newcommand{\heil}{He\,{\sc i}\,$\lambda$5876}
\newcommand{\heii}{He\,{\sc ii}}

\newcommand{\heiiol}{He\,{\sc ii}\,$\lambda$4686}
\newcommand{\mgii}{Mg\,{\sc ii}}

\newcommand{\nai}{Na\,{\sc i}}
\newcommand{\nail}{Na\,{\sc i}\,$\lambda\lambda$5890,5896}
\newcommand{\oiiiuv}{O\,{\sc iii}]}

\newcommand{\oiii}{[O\,{\sc iii}]}
\newcommand{\oiiil}{[O\,{\sc iii}]\,$\lambda$5007}

\newcommand{\oii}{[O\,{\sc ii}]}
\newcommand{\ssii}{[S\,{\sc ii}]}

\newcommand{\nii}{[N\,{\sc ii}]}

\newcommand{\fe}{Fe}
\newcommand{\feii}{Fe\,{\sc ii}}

\newcommand{\hst}{{\it HST}}

\newcommand{\ebv}{\ensuremath{\mbox{E(B-V)}}}

\newcommand{\kms}{\ensuremath{{\rm km~s}^{-1}}}
\newcommand{\kev}{\ensuremath{\mbox{keV}}}

\newcommand{\flux}{erg s$^{-1}$ cm$^{-2}$ }
\newcommand{\fluxl}{erg s$^{-1}$ cm$^{-2}$ \AA$^{-1}$}
\newcommand{\fluxhz}{erg s$^{-1}$ cm$^{-2}$ Hz$^{-1}$}

\newcommand{\Hoeq}{\ensuremath{H_0=75\,\kms \mbox{Mpc}^{-1}}}
\shorttitle{QSO UV-optical Spectral Properties}
\shortauthors{Shang et al.}

\begin{document}

\title{SPECTRAL PROPERTIES FROM \lya\ TO \ha\ \\
FOR AN ESSENTIALLY COMPLETE SAMPLE OF QUASARS I: Data}


\author{Zhaohui Shang,\altaffilmark{1,2}
Beverley J. Wills,\altaffilmark{3}
%\\
D. Wills,\altaffilmark{3}
Michael S. Brotherton\altaffilmark{2}
}

\altaffiltext{1}{Department of Physics, Tianjin Normal University,
Tianjin 300074, China. shang@uwyo.edu}
\altaffiltext{2}{Department of Physics and Astronomy, University of Wyoming,
Laramie, WY 82071, USA. mbrother@uwyo.edu}
\altaffiltext{3}{Department of Astronomy, University of Texas at
Austin, Austin, TX 78712. 
bev@astro.as.utexas.edu}

\begin{abstract}

We have obtained quasi-simultaneous ultraviolet-optical spectra for 22
out of 23 quasars in the complete PG-X-ray sample with redshift, {\it
z} $< 0.4$, and M$_{\rm B} < -23$.  The spectra cover rest-frame
wavelengths from at least \lya\ to \ha.  Here we provide a detailed
description of the data, including careful spectrophotometry and
redshift determination.  We also present direct measurements of the
continua, strong emission lines and features, including \lya,
\siivoivl, \civ, \ciii, \siiii, \mgii, \hb, \oiii, \heil+\nail, \ha, and
blended iron emission in the UV and optical.  The widths, asymmetries
and velocity shifts of profiles of strong emission lines show that
\civ\ and \lya\ are very different from \hb\ and \ha.
%The widths, asymmetries and velocity shifts of profiles of strong
%emission lines seem to put them into distinct groups --- while \civ\
%and \lya\ are very different from \hb\ and \ha, \mgii\ and \ciii\
%seems to be intermediate.  
This suggests that the motion of the broad line region is related to
the ionization structure, but the data appears not agree with the
radially stratified ionization structure supported by reverberation
mapping studies, and therefore suggest that outflows contribute
additional velocity components to the broad emission line profiles.

\end{abstract}



\section{INTRODUCTION}

QSOs appear to be signposts to galaxy evolution.  Supermassive black
holes have been discovered in nearby galaxies, with masses \mbh\
tightly related to host-galaxy bulge properties, e.g., the stellar
velocity dispersion \citep[][and references therein]{Trem02} and
luminosity \citep[][\& references therein]{MarHun03}.  There is little
doubt that these are the black-hole relics of the luminous QSOs in
their heyday at redshifts $z \approx 2$--3.  It is likely that QSOs
and their hosts evolve symbiotically.  The host supplies fuel to an
accreting black hole, perhaps through merger-driven star formation.
To enable fuel to feed the disk, the central system must lose angular
momentum, with this loss possibly via winds seen in emission
\citep{LeiMoo04,Leig04} and absorption or as more collimated jets.

%Sub-relativistic winds may play an 
%important role in limiting bulge formation \citep[e.g.,][]{Fab06}.
%Thus the black hole and host galaxy bulge evolve symbiotically, giving
%rise to tight coupling of black hole and bulge properties.

% Almost certainly, QSOs
% and their hosts evolve symbiotically.  The host supplies fuel to an
% accreting black hole, perhaps through merger-driven star formation.
% To enable fuel to feed the disk, the system must lose angular
% momentum, with this loss thought to be via winds seen in emission
% \citep{LeiMoo04,Leig04} and absorption
% %\footnote{http://www.nhn.ou.edu/$~$leighly/VImeeting/index.html}
% and as radio or X-ray jets.  Sub-relativistic winds may play an
% important role in limiting bulge formation \citep[e.g.,][]{Fab06}.
% Thus the black hole and host galaxy bulge evolve symbiotically, giving
% rise to tight coupling of black hole and bulge properties.

Fundamental parameters of the central engine are bolometric luminosity
\lbol\ ($\propto$ \ensuremath{\dot{M}}) representing the fueling rate
and efficiency, black hole mass \mbh, and Eddington accretion ratio
\lledd, also angular momentum of the black hole.  Observationally,
there are many trends and potential clues to the nature of this
central engine, from the radio to X-ray, from spectral energy
distributions, and lines in emission and absorption.  
%Many are summarized by an empirical picture put forth by Elvis (2000,
%2006).
While any understanding is incomplete without including the entire
observational domain, here we are concerned with QSOs' UV-optical
continua and emission lines.  Much of the diversity in QSO UV-optical
spectra can be accounted for by two strong empirical relationships
apparently related to
fundamental parameters of the central engine: 
Boroson \& Green's
eigenvector 1 relationship \citep[BGEV1,][hereafter BG92]{BG92}
and the Baldwin Effect (BE).

The original BE shows an anticorrelation between \civ\ equivalent
width (EW) and UV luminosity \citep{Bald77}.  It was later confirmed
that many other UV lines show similar relationships
\citep[e.g.,][]{Kinn90,Laor94a,Laor95,Will99a} which may depend on the
ionization potentials of the corresponding ions
\citep{EspAnd99,Gree01,Kura02,Croo02,Diet02}.  

BGEV1 is the first (i.e., most significant) eigenvector discovered
from a principal component analysis (PCA) of measured quasar
emission properties in the \hb\ region \citep{BG92}.  PCA is a
multivariate analysis.  It defines new eigenvectors, which are linear
combinations of input observables and can reveal important
relationships of the observables \citep[see][for a detailed
description]{FraWil99}.  The original Eigenvector 1
relationship found in the \hb\ region is characterized by the strong
anticorrelation between \oiii\ and \feii\ strengths, and involves
other parameters such as \hb\ FWHM and asymmetry (BG92).  BG92
suggested this set of relationships was driven by \lledd.  This was
further supported by \citet{L94, L97}, additionally finding that
narrower \hb, stronger \feii, and weaker \oiii\ corresponds to steeper
soft X-ray spectra.

In an attempt to understand the above relationships, we obtained
quasi-simultaneous UV-optical spectra over the entire \lya--\ha\ range
for 22 of the 23 quasars of the Bright Quasar Survey investigated for
soft X-ray properties by \citet{L94,L97} (see \S\ref{sec:obs}).  We
extended the spectra to the UV and showed, by principal components
analysis, the important extension of BGEV1 into the ultraviolet, as
well as demonstrating that BG's luminosity-dependent eigenvector was
actually directly related to the Baldwin Effect in the UV
\citep{Will99b,Will99c,FraWil99,Will99a,Wil00}.  \citet{Shan03}
carried out a spectral principal components analysis (SPCA) of the
same sample.  SPCA decomposes the input spectra into fewer significant
orthogonal (i.e., independent) principal eigenvectors (or
eigenspectra), which reveal important relationships among the spectral
features \citep[see][]{Fran92,FraWil99}.  The first principal eigenvector in
\citet{Shan03} represents the BE, and nicely links the equivalent
widths of many ultraviolet lines with the optical, including the broad
\heiiol\ line, whose inverse luminosity dependence was directly
demonstrated by \citet{BG92} and \citet{Boro02}.  The second principal
component was the result of differences in continuum shape.  The third
principal component extends BGEV1 to the UV, showing correlations
among broad line widths, and strengths.  These first three principal
components accounted for 78\% of intrinsic variance among spectra in
the sample, demonstrating that these relationships are so clear and
obvious, that only a small sample is needed to reveal them.

The technique of SPCA has the distinct advantage that correlations
among  emission and absorption lines can be seen independently of
having to define a continuum for these blended features, or to measure
specific line-profile parameters.  Results over the entire wavelength
range are immediately visible, revealing correlations that may not
otherwise have been sought.  However, disadvantages of SPCA are that
the only directly interpretable relationships are those that can be
represented by linear relations among the  many flux bins along the
spectrum.  Non-linear dependences caused by differences in line
widths, asymmetries, and shifts, or non-linear relationships among
emission line EWs, as would be present in the BE over a sufficiently
wide luminosity range, are more easily investigated by direct line
measurements.

Thus we must also investigate relationships among directly measured
spectral parameters.  Here we provide a detailed description of the
data and direct measurements of the continua, and all strong emission
lines and features.  We also present the distribution and
relationships of emission line profile properties, including line
width, velocity shift, and asymmetry.  We will expand our previous
analyses with new measurements and more parameters in a subsequent
paper.  For our cosmology, we choose zero cosmological constant,
\Hoeq, and $q_0=0.5$.

%%%%%%%%%%%%%%%%%%%%%%%%%%%%%%%%%%%%%%%%%%%%%%%%%%%%%%%%%%%%%
\section{SAMPLE AND DATA}

%%%%%%%%%%%%%%%%%%%%%%%%%%%%%
\subsection{Sample \label{sec:sample}}

We use the same complete sample of 23 QSOs\footnote {PG~1048+342 was
not observed in the UV band for non-scientific reasons
(Appendix~\ref{ap:data}), so this omission does not bias the sample,
and we have used 22 QSOs in our UV-optical analyses.  } selected by
\citet{L94,L97} from the Bright Quasar Survey
\citep[BQS;][]{SchGre83}.  Laor et al. aimed to test models of the
optical to X-ray continuum, and photoionization of the regions
producing optical emission lines.  They therefore optimized the sample
for observations of the rest-frame soft X-rays by choosing a complete
sample of bright, low redshift ($z\leq0.4$) QSOs, with 
%H\,I 21\,cm 
Galactic \hi\ column density $N_{\rm HI} < 1.9\times 10^{20} {\rm cm}
^{-2}$ to minimize soft-X-ray absorption.  Note that the
incompleteness of the BQS discussed by  \citet[][]{Jes05} does not
bias our low-redshift ($z<0.4$) sample, since the color-related bias
only affect objects with redshifts of $0.5 < z < 1.0$.  For our
UV-optical observations, the low redshift allows investigation of UV
spectra with much reduced contamination from intergalactic absorption
lines and the low \nhi\ ensures small corrections for Galactic
reddening.  
%[** The following could be reserved for the next paper:\\
%Optimization of the sample for rest-frame soft-X-ray observations
%allows comparison of the optical-UV continuum and emission lines with
%continuum ionizing flux.]
The QSOs in our sample are listed in Table~\ref{tb:sample} along with
their redshifts, magnitudes, soft X-ray spectral index \alphax, and
radio loudness.  

%%%%%%%%%%%%%%%%%%%%%%%%%%%%%%%%%%%%%%%%%%%%%%%%%%%%%%%%%%%%%%%%%%%%%%%%%%%%
\subsection{Observations\label{sec:obs}}

Figure~\ref{fg:wavecover} presents the wavelength coverage for all
spectral observations and the colors distinguish different observing
runs.  Table~\ref{tb:log} lists observing dates for each spectrum. 

UV spectra of all 22 objects were obtained in ACCUM mode with {\em
Hubble Space Telescope} (\hst) Faint Object Spectrograph \citep[FOS,
][]{Key95}, covering wavelengths from below \lya\ to beyond the
atmospheric cutoff near 3200\,\AA\ in the observed frame.
Instrumental resolution for all the UV data is equivalent to
$\sim$230\,\kms\ (FWHM).  

Optical data were obtained at McDonald Observatory except for
some \hst\ FOS data and a little early archival data (see
Appendix~\ref{ap:data}).  

On the Harlan J. Smith 2.7m reflector, the 
Large Cassegrain Spectrograph (LCS) was used with a
Craf-Cassini detector (CC1, a thick chip with excellent cosmetic quality).
Some observations of 1996 April and May used Electronic Spectrograph 2 (ES2) on
the Otto Struve 2.1m reflector (F1) with CC1 and a thinned  Texas Instruments
detector (TI1), respectively.  

Optical observations were made generally at airmasses less than 1.3,
using long slits -- both narrow and wide.  The narrow slit (1\arcsec
-- 2\arcsec) was used for best spectral resolution, and to reduce
scattered, background sky, and host-galaxy light. Wide slits
(8\arcsec\ on LCS, 9.1\arcsec\ on ES2) were used for absolute
flux-density calibration (\S\ref{sec:red:opt}).  The wavelength
resolution with the narrow LCS slit and ES2 slit was typically $\sim
7.5$\AA\ FWHM, but 6.7\AA\ for ES2 in 1996 May, 
%(2.7\AA/pixel for LCS,
%2.64\AA/pixel for CC1 on ES2, and 3.35\AA/pixel for TI1 on ES2),
equivalent to 450 to $\sim$~300 \kms (FWHM) in the \hb\ to \ha\
region.
The slit orientation was east-west, except in one case where the
spectrograph was rotated to avoid a contaminating star.  Flux density
standard stars were observed several times each night, chosen to be
close in time and airmass to the QSOs.  These standards were chosen
from the \hst\ list \citep{Bohl96,Bohl00} and their flux calibration
files were obtained from \hst\ CALSPEC
(\url{ftp://ftp.stsci.edu/cdbs/cdbs2/calspec/}, February, 1996
update), to ensure consistent calibration between UV and optical data.
Preference was given to the available standards of highest priority
\citep[see][]{Bohl96,Bohl00}.  

To reduce the uncertainty caused by QSO intrinsic variability, we
attempted to get quasi-simultaneous optical observations.  These were
usually obtained within a month of the new \hst\ observations
for 11 objects and 2--10 months for another 5 objects.  
Quasi-simultaneous observations were not possible for HST archival
data.  In table~\ref{tb:log}, time gaps between observations of each object are 
listed for all the spectra, as well as actual observing dates.

A set of higher resolution spectra\footnote{Unpublished 
observations, Bingrong Xie, University of Texas at Austin.}
($\sim180$\,\kms\ FWHM) of the \feii(opt) --\hb --\oiii\ region were
obtained with the same spectrograph and telescope at McDonald
Observatory.  These spectra were used here only to determine the redshifts of most
of the QSOs in our sample with sufficiently strong \oiii\ lines
(\S~\ref{sec:redshift}).  These relatively high-luminosity AGNs have
broad enough intrinsic emission lines that there was no special
advantage to using the higher-resolution data for other analyses.

%%%%%%%%%%%%%%%%%%%%%%%%%%%%%%%%%%%%%%%%%%%%%%%%%%%%%%%%%%%%%%%%%%%%%%%%%%%%
\subsection{Data Reduction}

\subsubsection{\hst\ Spectra \label{sec:red:uv}}

UV \hst\ spectra were reduced and calibrated through the standard FOS
pipeline, 
%[** say which version of the pipeline - or date of analysis], 
which is described at \url{http://www.stecf.org/poa/FOS/index.html}.
The ACCUM mode observations were obtained in a series of 2- or
4-minute integrations.  The resulting light curves for each object
were checked for possible pointing problems.  The light curves are
usually flat, indicating stable pointing and tracking.  Occasionally a
decline in the light curve indicated a loss of signal, so integrations
were scaled to match the ones with the highest flux, which are likely
to be correct.  The validity of this is proved later because flux
density levels for spectra taken with different gratings agree well in
the regions of overlap.

Wavelength scales have been calibrated within the FOS pipeline, which uses vacuum
wavelengths.  Post Operational \hst\ Archives (POA) show that the systematic uncertainty
of the wavelength zeropoint for FOS/BLUE spectra is about $\pm 1$\,\AA\ ($\sim$230~\kms)
\citep{Rosa01}.  No effort has been made to correct this for this
sample.  More information can be found at \url{http://www.stecf.org/poa/FOS/}

%The small dip resulting from the flat fielding residual (observed
%wavelength 1950\,\AA) occurs at different rest wavelengths for
%different objects.  This does not cause any problem in later
%analysis, because the feature is weak compared with strong emission
%lines, and it is narrow if in the continuum region.

\subsubsection{Ground-based Spectra \label{sec:red:opt}}

Standard packages in IRAF were used to reduce the optical data.
Obvious cosmic ray features were removed from the images by hand.
After extraction, the spectra were checked again for cosmic ray
features that were also removed. 

For narrow slit spectra (1\arcsec--2\arcsec), variance-weighted
(optimal) extraction was used to achieve the best S/N ratio with an
aperture size of $\sim$10 pixels 
(usually the seeing $<$ 3 pixels FWHM).  For wide
slit spectra (8\arcsec, for flux calibration), boxcar extraction was
used and the aperture size is set as long as 26 pixels ($\sim13''$)
to include as much light from the object as possible.  The background
regions were defined on both sides of the extraction aperture with a
buffer region.  The background was fitted with a low-order polynomial
and subtracted during the extraction.

Wavelength scales were calibrated using neon and argon lamp spectra.
This wavelength scale was checked against wavelengths of night sky
lines.  For some spectra night sky lines were used to apply a
zero-point wavelength correction to the wavelength scale.

Absolute flux calibration was achieved by using standard star spectra
observed on the same night.  Differential atmospheric extinction was corrected
with extinction files suitable for McDonald Observatory in different
seasons (Barker, E., private communication).  As the standards were
observed close in airmass to the QSOs, extinction was a minor correction.

Atmospheric absorption bands 
include the very strong O$_2$ $A$ (7606\,\AA) and $B$
(6871\,\AA) bands, H$_2$O $a$ band ($\sim$8200\,\AA) and weaker bands
between $\sim 8000-10000$\,\AA.  In order to correct for these, we
create a correction spectrum for each QSO spectrum using a hot star
spectrum (usually a standard star) obtained on the same night, close
in time and airmass.  

Finally, slit loss for the narrow-slit QSO spectra is corrected by
scaling them to match the shape and absolute flux-density level of the
wide-slit QSO spectra, which are assumed to include all the light from
the QSO.  This is done by creating a ratio spectrum of the wide-slit
(8\arcsec) spectrum and the narrow-slit (1\arcsec\ or 2 \arcsec)
spectrum, fitting its continuum with a low-order smooth curve, and
using this curve to correct the narrow-slit QSO spectrum.  In
principle, the scaling is different at different wavelengths, but
usually it is very close to a constant over the wavelength range 
($< 3000$\,\AA) of a single spectrum.  For some cases, a slight (smoothed)
slope has to be applied in order to match the wide-slit spectra.

The slit-loss correction may increase the uncertainty of flux caused
by the host-galaxy contamination because the wide-slit spectra collect
more host galaxy light than the narrow-slit spectra.  However, all
objects in our sample are luminous QSOs and their host galaxies are
very faint.  During the spectral extraction, the background fitting
and subtraction have removed most of the host galaxy contribution.
Moreover, host galaxies for luminous QSOs are usually elliptical
galaxies \citep[e.g.,][]{McLRie94a,McLRie94b,Bahc97,Dunl03}, and their
contamination to the QSO spectra is mostly in the red wavelengths
beyond $\sim$4000\AA.  We estimate that the host galaxy contamination
in the narrow-slit spectra is usually $<5\%$.  After the slit-loss
correction, the contamination is still $<10\%$ in general, and
$<20\%$ for the worst case (PG1115+407) based on the
host galaxy photometry \citep{McLRie94a}.  Finally, this may 
affect our line measurement of only the \ha\ region.

%%%%%%%%%%%%%%%%%%%%%%%%%%%%%%%%%
%\subsubsection{Combined UV-optical Spectra sec:red:com\label{sec:red:com}}
\subsubsection{Combined UV-optical Spectra \label{sec:red:com}}

The ground-based spectra were transformed to vacuum wavelength
scales to be consistent with those from \hst.  UV and optical spectra
for each object were then combined in the observed frame.
When the overlap region is large, the spectra were combined with
different weights which are reciprocals of their variance calculated
for a common clean continuum region within the overlap region.  

In general, the flux-densities in the overlapping spectral regions agree 
within a few percent, often to $\pm 1$\%.  Sometimes differences are large -
attributable to poor calibration or, more likely, to intrinsic QSO variability
for non-quasi-simultaneous observations.
So, when combining spectra for each object, we chose, as a reference, a
spectrum with the best flux-density calibration based on, for example,
observing conditions, agreement among standard star observations on the same 
night, and consistency of repeated observations and among the well-calibrated 
spectra in the overlap regions.  This is fairly easy because there are always 
several spectra for one object and agreement often exists for a few spectra.  All
spectra are scaled, if necessary, to match the flux of the reference
spectrum, and the scaling factors for individual spectra are listed in
Table~\ref{tb:log}.
For a given object the relative flux-density calibration over the whole
wavelength range should be $<5$\%.  Considering the good flux density
calibration of the reference spectrum for each object, the accuracy of absolute
flux density calibration should approach that of the standards defining
the system ($\sim$3\% over most of the wavelength range,
\citet{Bohl96}).
Host galaxy contamination could add some additional uncertainty for 
wavelengths longer than 4000\AA\ (\S\ref{sec:red:opt}).

After combining spectra, strong geocoronal lines at (observed frame)
1215\,\AA, 1302\,\AA, and 1355\,\AA\ were removed from the spectra.
Strong interstellar medium (ISM) absorption lines were identified using
their expected wavelengths and removed by hand.  
% [Note:  At one stage I measured waveengths of interstellar lines,
% relative to 21cm, for wavelength calibration.] [This was
% incorporated in a study done by F.Kerber of ESA.] 

Galactic reddening was removed with an empirical mean extinction law
\citep*{Card89}, assuming $R_V = A_V/E(\bv) = 3.1$, a typical value
for diffuse interstellar medium.  $E(\bv)$ is obtained from NED
(Table~\ref{tb:sample}) based on the dust maps created by
\citet{Schl98}.

%\subsubsection{Redshifts sec:redshift \label{sec:redshift}}
\subsubsection{Redshifts \label{sec:redshift}}

We assume that the narrow emission line region in AGN is at the
systemic redshift.  We note that there are studies reporting
blueshifted \oiii\ based on the assumption that \hb\ is at the
systemic redshift \citep[e.g.,][]{Zama02,Sule00}.  A recent study
\citep{Boro05}, using more reliable lower ionization lines \oii, \nii,
and \ssii\ as the reference, shows that the average blueshift of
\oiii\ is 40\,\kms, with 5\% of AGN having blueshifts larger than 100\,\kms.
For our sample, we use \oiii\ 5006.8\,\AA\ to define the rest frame.

For most objects in our sample, we have a separate set of higher resolution 
spectra (\S\ref{sec:obs}) that we used to measure \oiii\ redshifts.
After subtraction of optical \feii\
emission blends using the same \feii\ template and method used by
BG92, the redshift is measured using the \oiii\ $\lambda$5006.8 line
(Table~\ref{tb:sample}).  Redshifts for three objects with measurable
\oiii\ but without the higher resolution spectra, PG1001+054, PG1425+267 and 
PG1512+370, were measured in the same way from the lower resolution spectra.
The measurement uncertainty of redshifts using \oiii\ is 0.0002.
For objects with very weak \oiii\ emission (PG1402+261, PG1444+407,
and PG1543+489), other emission lines, such as \hb\ and deblended
\ciii, were used and a less accurate redshift (uncertainty $\sim
0.001$) was assigned.  The spectra are presented on a rest-frame
wavelength scale in Figure~\ref{fg:spec}.  The flux density is in the
observed frame.

%%%%%%%%%%%%%%%%%%%%%%%%%%%%%%%%%%%
%\section{Spectral Measurements sec:measure\label{sec:measure}}
\section{Spectral Measurements \label{sec:measure}}

\subsection{Continuum and Spectral Indices \label{sec:measure:index}}

We use power-law spectral indices to characterize the QSO continua.
Unless noted, the power-law indices we use through this paper are all
$\alambda$ ($\flambda \propto \lambda^{\alambda}$) except for soft
X-ray spectral index \ax\ and soft X-ray to optical spectral index
\aox, which are \anu\ ($\fnu \propto \nu^{\anu}$).  It is easy to
convert between $\alambda$ and $\anu$  using $\alambda+\anu=-2$.  

It is customary to measure the UV-optical spectral index by fitting
the continuum with a power-law, but in many objects a single power-law
cannot fit the entire region.  This was noticed before in a similar
sample \citep{Shan05}.  In radio-quiet QSOs, a UV downturn is often
seen \citep[e.g.][]{OBri88,SuMa89}, and is present in many composite
spectra \citep[e.g.][\& references therein]{Vand01}.  In the spectra
of radio core-dominant QSOs, an additional synchrotron component may
be present \citep[e.g.][]{MaMo86,Wil91}.  Line- and edge-free regions are hard
to find in the QSO spectra because of the many broad emission lines
and blends.  In fact, when \citet{Laor97b} studied the narrow-line
($\sim 900\,\kms$) quasar I~Zw~1 with high S/N ratio, and
high-resolution spectra, weak emission lines and blends (e.g., \feii)
were found virtually everywhere.  There are essentially no
emission-line-free regions in QSO spectra.  The continuum windows are
only relatively clean regions and can be different from object to
object.  We therefore choose six common narrow continuum windows where
the emission lines appear very weak: 1144--1157\AA, 1348--1358\AA,
4200--4230\AA, 5600--5648\AA, 6198--6215\AA, and 6820--6920\AA.

As a useful measure of the overall continuum slopes we define a
UV-optical spectral index \auvo\ in the range $\sim$1200--5500\AA\ and
a red optical spectral index \aored\ in the range $\sim$5500--8000\AA\
by fitting a power-law to a pair of selected continuum windows.   We
try to keep the emission features above the fitted power-laws in
the corresponding regions.  It seems that the continuum window
5600--5648\AA\ is needed for both power-laws for most objects.  The
continuum windows used for each object and the fitted power-law
indices are listed in Table~\ref{tb:cont} and marked in
Figure~\ref{fg:spec}.  
%Since the continuum windows are very narrow, this fitting process is
%more like defining each power-law with two points.  The power-laws
%cannot be treated as the true continua of the spectra, but the
%spectral indices provide information on the overall continuum slopes.

If the difference between the two spectral indices is less than 0.20,
we consider that a single power-law is a good approximation to the
continuum over the whole wavelength range.  Six out of 22 objects meet
this criterion (see Table~\ref{tb:index}).  This can also be visually
checked in Figure~\ref{fg:spec}.

We have also obtained the measurements in the soft X-ray region for this sample
\citep{L97} and calculated the optical to soft X-ray spectral index
\aox\ between 2500\AA\ and 2\,keV.  The 2500\AA\ flux is estimated
using the fitted power-law continuum.  The \aox\ is listed in
Table~\ref{tb:index}.

%%%%%%%%%%%%%%%%%%%
\subsection{Emission Lines \label{sec:measure:line}}

For each of several emission-line regions, we fit simultaneously a
local power-law continuum and different emission-line components,
using $\chi^2$ minimization within the IRAF package {\em specfit}
\citep{Kris94}.  Figure~\ref{fg:fitline} shows examples of the fitting
in different regions.  We used a narrow and a broad Gaussian component
to fit each strong, broad emission line.  A velocity shift is allowed
between the two components to account for the asymmetry of the line
profile.  
%The narrower component mostly accounts for the line-core of broad
%lines, not the NLR emission which is not obvious in most cases.
For \ha\ and \hb, we have also included a third Gaussian component to
account for the obvious NLR emission on top of the broad-line profile
for several objects.  For weak or narrow emission lines, one Gaussian
component is used in the fitting.

We used symmetric Gaussian profiles, so each profile has three
parameters: flux, width, and central wavelength.  In order to 
avoid too many free parameters, for each emission-line
region we assume the same width for similar components of
different lines, tie together the
wavelengths of some lines based on their laboratory wavelengths, and
assume line intensity ratios for some lines based on their statistical
weights.  We list in Table~\ref{tb:fitpar} all the free and dependent
parameters in each region and how the parameters are related.  For
example, in the \hb\ region, \oiii\,$\lambda$4959 has the same width and 1/3 of
the flux of \oiii\,$\lambda$5007; its wavelength is tied to that of 
\oiii\,$\lambda$5007 \citep[see also][]{Kris94}.

We have included \fe\ templates to account for the complicated \fe\
emission only in the optical \hb\ region and UV \mgii\ regions.  Both
the optical \feii\ template \citep{BG92} and the UV \fe\ template
\citep{VesWil01} were derived from the spectra of the narrow-line
quasar I~Zw~1.  These templates were allowed to vary in both intensity
and broadening to match the \fe\ emission in different objects, and
they are fitted simultaneously with the local power-law continuum and
emission-line components in {\em specfit}.  
In doing so, we assume that the \fe\ emission is above a power-law
continuum, same assumption as these templates were derived.
Iron can also be important
in and near the red wing of \civ\ and introduces uncertainty in the
related measurements for \civ.  We discuss this in
\S\ref{sec:measure:error}.  

Figure~\ref{fg:fitline} shows examples of our fitting results and
individual components in each region.  We also show our treatment of
broad absorption lines or deep absorption features, in the lower right
of the figure.  There is obvious strong absorption of \lya, \civ, or
\siiv\ in three objects (PG1001+054, PG1114+445, PG1411+442).  We
attempted to fit the affected regions in  two ways.  First, we
excluded the absorbed regions before fitting; second, we fit the
region, including additional absorption Gaussian profiles at the
absorption positions.  The two methods give consistent results
and we use the results from the first method
in the following analyses.

Our fitting process separates the blended lines, but with possible
large uncertainties because of uncertainties in continuum fitting and
our assumptions about the relationships between different lines
(Table~\ref{tb:fitpar}), e.g., same width for two lines.  This affects
the weak lines the most and their fluxes are less reliable.  
They are used only to understand and remove their contribution on the
strong lines.  Even for some strong lines, although we define
different emission lines in the region, we cannot really deblend them
without arbitrary assumptions.  Therefore, we present some
measurements of just the total blend, such as \siivoivl\ and
\nai+\heil.

For all individual broad lines other than \hb\ and \ha, we use two
Gaussian profiles at most and do not include an NLR component.  This
sometimes cannot fit the narrow peak for some lines.  These peaks
could indicate an NLR component, but without other supporting
evidence, we cannot identify or separate them from the broad
components.  In fact, the lack of a \hb\ NLR-like (or \oiii-like) peak
for \civ\ has led to a commonly accepted view that the NLR component of
\civ\ is very weak, if not absent \citep[e.g.,][]{Will93,VesPet06},
although some studies \citep[e.g.,][]{Sule99} claim a relatively
strong \civ\ NLR component by allowing a fitted component generally
much broader than \oiii.  Whether those residual peaks in our fitting
are a NLR component or not, they contribute very little to the total
flux of the strong emission lines.  
The FWHM might have been overestimated a little if there is a 
    residual peak but no real NLR emission, but the uncertainty is
    still within the error of the measurements (\S\ref{sec:measure:error}).
Our overall fit for each region is
acceptable in terms of deblending and obtaining profile parameters of
strong emission lines.  These fits can be assessed by inspection of
Figure~\ref{fg:fitline}.

Emission-line flux, EW and profile parameters are derived from the
fitting results.  For an emission line with a single Gaussian profile,
the calculations are straightforward.  For a strong broad emission
line with two Gaussian profiles in the fitting, we add the two
profiles and local continuum to form a  model spectrum of the emission
line region (excluding an NLR component) and then derive the
parameters from the resulting model. EW is calculated using the line
flux and the fitted local continuum at the emission line wavelength
and then transformed to the rest-frame.  All the line velocity shifts
are for the peak of the model line profile and relative to the
systemic redshift.  FWHM and asymmetry are also derived from the
emission line model.  We define a line asymmetry parameter
\citep{DeRo85,BG92,CorBor96}, 
\[ \mbox{Asymm} =
\frac{\lambda_c(\frac{3}{4}) - \lambda_c(\frac{1}{4})} {\mbox{FWHM}},
\] where $\lambda_c(\frac{3}{4})$ and $\lambda_c(\frac{1}{4})$ are the
wavelength centers of cuts at $\frac{3}{4}$ and $\frac{1}{4}$ of the
line peak flux density.  A positive value indicates excess light in
the blue wing of an emission line.

We provide the measurements of all strong emission lines in
Table~\ref{tb:stronglinesA}--\ref{tb:stronglinesE}.  The uncertainty
of the measurements is discussed in \S\ref{sec:measure:error}.

%%%%%%%%%%%%%%%%%%%%%%%%%%%%%%%%%%%%%%%%%%%%%%%%%%%%%%%%%%%%%%%%%%%%%%%%%
\subsection{Small Blue Bump \label{sec:measure:sbb}}

%While we have specifically used the Fe templates to remove their
%effects on the strong emission lines, here we consider how to estimate
%their strength.

AGN UV-optical continuum spectra usually show a bump between
2000\,\AA\ and 4000\,\AA.  This is referred as the ``Small Blue Bump''
(SBB) and consists mainly of unresolved \fe\ blends and Balmer
continuum.  Although we include the \fe\ template in fitting the
\mgii\ region, the UV \fe\ emission estimated is not complete due to
the cutoff of the template at $\sim$3000\AA.  About 1/3 of the SBB
above 3000\AA\ is missing in the template, and the local continuum for
\mgii\ can also be affected in the fitting.  
We therefore attempt to measure the SBB
directly by integrating the spectrum between 2220 and 4010\,\AA\ above
a continuum and then subtracting the contribution of \mgii\ using its
fitting results.

We have tried different continua.  The first one is the global
UV-optical continuum defined by \auvo, but it is likely the SBB flux
is overestimated (Fig.~\ref{fg:fitsbb}) because the continuum is
defined over a wider wavelength range.
We then define a local
power-law continuum by connecting the spectrum between 2220 and
4010\AA.  In many cases, this underestimates the SBB.  Although either
can consistently give us a good estimate of the SBB for the whole
sample and both have a large uncertainty, the estimates are simply
related and both can roughly represent the SBB.  More likely, they can
be treated as the upper and lower limits of the SBB flux
(Table~\ref{tb:stronglinesD}).

% local cont. 2220-4010
% global cont. auvo
%%%%%%%%%%%%%%%%%%%

\subsection{Measurement Uncertainties \label{sec:measure:error}}

Since all our spectra are have high S/N ratio, the formal error in
the fitting process is not significant.
The major error comes from the
placement of the local continuum in each emission-line region.
Therefore, in order to estimate the uncertainties of the measured
quantities, we adopt a method similar to \citet{L94,L97}.  We change
the best-fit local continuum level by $1\sigma$ error at both ends of
the local emission-line fitting region, build new model line profiles,
and re-calculate the quantities (\S\ref{sec:measure:line}).
We start from the best-fit local continuum fluxes $F_{left}$ and
$F_{right}$ at both ends of a local continuum region.  The $1\sigma$
errors of the continuum fluxes are measured at these two wavelengths
as $\sigma_{left}$ and $\sigma_{right}$, respectively.  Between
$F_{left}\pm\sigma_{left}$ and $F_{right}\pm\sigma_{right}$, we have 4
combinations of the new local continuum.  We repeat the above
calculations 4 times, compare with the best-fit results, and pick up
the largest differences as the errors for each quantity.  For EWs,
corresponding new continua are used in the calculation.  The errors
are listed in Table~\ref{tb:stronglinesA}--\ref{tb:stronglinesE}.

The errors calculated for line velocity shifts \dv\ are always very
small ($<5\,\kms$), but the real uncertainty for velocity shift
results from the uncertainty of the redshifts defined and measured
using \oiiil\ (\S\ref{sec:redshift}).  The redshift measurement
uncertainty can be roughly evaluated from the velocity shift of
\oiiil\ (Table~\ref{tb:stronglinesC}), which is supposed to be zero.

For spectral indices, we use the same method to estimate their
uncertainties.  We note that another possibly important source of
error for \auvo\ comes from the Galactic reddening correction
(\S\ref{sec:red:com}).  We do not attempt to estimate this error
because of the lack of errors of \ebv\ for individual objects.

Blended iron emission causes complication in measuring other emission
lines.  We have used \fe\ templates to fit and remove the \fe\
emissions in the \hb\ and \mgii\ regions where \fe\ is strong, but we
do not include the \fe\ template in the \ciii\ and \civ\ regions,
where \fe\ can sometimes be important.  Assuming the iron intensities
in the \mgii\ and \civ\ regions scale together, and using the fitting
results from the \mgii\ region and the same UV \fe\ template, we
estimate the \fe\ contribution within $3\sigma_{broad}$ around the
\civ\ line center, where $3\sigma_{broad}$ is the Gaussian $\sigma$ of
the \civ\ broad Gaussian component, which mainly models the line
wings.  We found that the \fe\ is only $1.9\%$ of the total \civ\ flux
on average, and $4.1\%$ for the worst case.  In fact, in the fitting
process of the \civ\ region, part of the \fe\ in the \civ\ red wing
has been treated as pseudo-continuum and is not included in the
measured \civ\ flux.  Therefore, the uncertainty in \civ\ flux caused
by \fe\ should be less than $4\%$ for all objects.

%%%%%%%%%%%%%%%%%%%%%%%%%%%%%%%%%%%%%%%%%%%%%%%
\section{Estimation of Physical Parameters}

% [Should one use FWHM rather than sigma that the latest Peterson et al.
% papers show is an improvement?]

We have estimated the black hole mass and accretion rate using the
empirical method developed from reverberation mapping studies
\citep{Kasp00,Pete04}. 
Assuming virial motion of the BLR,
\begin{equation}
\label{eqMbh}
M_{BH} = R_{BLR}\ v^2/G.
\end{equation} 
where velocity dispersion $v$ is estimated from 
\hb\ line width, $v=\sqrt{3}/2\ \mbox{FWHM}(\hb)$, 
and the size of the BLR $R_{BLR}$ has an empirical
relationship with luminosity \citep{Kasp00},
\begin{equation}
\label{eqRblr}
 R_{BLR} = 32.9^{+2.0}_{-1.9} \left[ \frac{\lambda L_\lambda (5100 \mbox{\AA})}
{10^{44} \mbox{erg s}^{-1} } \right]^{0.70\pm0.033} \mbox{  light days}.
\end{equation}
$\lambda L_\lambda (5100 \mbox{\AA})$ is calculated using the fitted 
local continuum in the \hb\
region and our adopted cosmology  ($\Lambda=0,\ \Hoeq, q_0=0.5$).
Similarly, we also obtain a UV luminosity \llciv=$\lambda L_\lambda
(1549 \mbox{\AA})$, using the local
continuum in the \civ\ region at 1549\,\AA.

Following \citet{Kasp00}, we define the bolometric luminosity
\lbol=9\,\lllambda(5100\mbox{\AA}).  Knowing the black hole mass, we
can then estimate the Eddington accretion ratio, \lledd, where $\ledd
= 1.25 \times 10^{38} (\mbh/\msun)$.  The bolometric luminosity
defined here is only an estimate of the true bolometric luminosity.
Different scaling factors for monochromatic luminosity exist, e.g.,
\lbol$\approx$13.2\,\lllambda(5400\mbox{\AA}) \citep{Elvi94}.
\citet{Shan05} estimated \lbol\ for a sample with good far-UV-optical
spectra and found \lbol $\approx1.5\times$
9\,\lllambda(5100\mbox{\AA}).  Using photoionization modeling to
determine the ionizing continuum, \citet{Wand99} deduce \lbol
$\approx$10\,\lllambda(5100\mbox{\AA}).  Recently, using
multi-wavelength data, \citet{Richa06} show
\lbol=($10.3\pm2.1)$\,\lllambda(5100\mbox{\AA}), and point out that
deriving a bolometric luminosity from a single optical luminosity can
lead to errors as large as 50\%.  We note  that our adopted \lbol\ may
have large uncertainty, but the relative error within the sample is
smaller.  We list these derived parameters in Table~\ref{tb:mbh}.

%%%%%%%%%%%%%%%%%%%%%%%%%%%%%%%%%%%%%%%%%%%%%%%
%%%%%%%%%%%%%%%%%%%%%%%%%%%%%%%%%%%%%%%%%%%%%%%
%%%%%%%%%%%%%%%%%%%%%%%%%%%%%%%%%%%%%%%%%%%%%%%
%\newpage

%\section{Spectral Properties and Correlation Analysis \label{sec:corr}}

\section{Properties of Emission Line Profiles \label{sec:fwhm} \label{sec:shift}
\label{sec:asymm}}

We compare here the profile properties of lines arising from different
atomic transitions, and present the correlations of their FWHM,
velocity shift, and asymmetry parameter.  We leave more detailed
analyses to Paper~II \citep{Will07}.  We denote the Pearson
correlation coefficient by $r$ and the two-tailed probability of a
correlation arising by chance, by $p$.  Note that we have previously
discussed the line-width correlations based on different line
measurements, in \citet{Wil00}.

\subsection{Line Width}

As illustrated clearly in Figure~\ref{fg:fwhm},  \ha\ FWHM has the
strongest correlation with \hb\ FWHM.  The scatter of the correlations
with \hb\ FWHM gets larger for \mgii\ and \ciii, and the correlation
virtually disappears for \lya\ and \civ.  However, a correlation
between \lya\ and \civ\ FWHMs exists ($r=0.81,p<10^{-5}$).  Similar
results have been noticed before, e.g., \citet{CorBor96} also found
that the FWHMs of \civ\ and \lya\ are more strongly correlated with
each other than with \hb.  In QSOs' photoionized regions the \lya\
behavior is expected because a significant fraction is emitted from
the high-ionization region along with CIV.  Therefore we treat \lya\ as
a high-ionization line.

We also notice that \civ\ and \lya\ are not necessarily broader than
\hb.  This seems to argue against a simple radially stratified
ionization structure of BLR as suggested by reverberation mapping
studies \citep[e.g.,][]{Pete91,Kori95}, but it is more likely that
the total line width is affected by a wind component.  
We discuss this more in
\S\ref{sec:profile}.

We note the possible problem with \lya\ measurements here.  When
fitting \lya\ region, we assume that \nv\ has the same profile as
\lya\ and tie its wavelength to \lya.  If the assumptions are wrong,
it will affect the measurements of \lya.  The \lya\ line velocity
shift is affected little because it is measured from the \lya\ line
peak, however, the line width can be affected when \nv\ is strong and
the asymmetry parameter can be affected if \nv\ is not correctly
de-coupled from the \lya\ red wing.  Without knowing the true profile
of \nv, this situation remains true for any attempt of de-blending
\lya\ and \nv\ under assumptions.

\subsection{Line Peak Shift}

We plot the distributions of emission-line velocity shifts in
Figure~\ref{fg:shift} and list their statistics in
Table~\ref{tb:shift}.  It is clear that \civ\ shows significant
blueshifts.   This agrees with the results from a large sample in
\citet{BasLao05} in general, although the shift parameter is defined
in the units of FWHM there.  There is evidence that \lya\ and \mgii\
are also blueshifted, but not as much as \civ.  \ha\ and \hb\ show
small redshifts, but it seems there is not a preferred direction for
\ciii\ velocity shift, suggesting that \ciii\ may also be good for
defining the QSO redshift if narrow lines are not available.  However,
we note that the wavelength of \ciii\ can be measured accurately
enough only when the broad emission lines in this spectral region are
sufficiently narrow to allow a decomposition.  Although the dispersion
of line shifts is large, it seems, from blueshift to redshift, that a
sequence is formed for \civ, \lya, \mgii, \ciii, \hb, and \ha,
suggesting an ionization level dependence.  This agrees with some
earlier studies \citep[e.g.,][]{Gask82,Corb90,TytFan92}, but not
others \citep[e.g.,][]{Laor95}.

Our correlation analyses (Table~\ref{tb:corrshift}) further show that
the shifts of \civ\ and \lya\ are correlated ($r=0.81, p=4.9\times
10^{-6}$), and those of \ha\ and \hb\ are also correlated ($r=0.62,
p=0.002$), but the shifts of \civ\ and \lya\ do not seem to be related
to that of \ha\ or \hb.  In terms of the correlation coefficients,
\mgii\ and \ciii\ seems to be related to \ha\ and \hb\ more closely
than to \civ\ and \lya.   

\subsection{Asymmetry }

We have measurements of asymmetry for only four emission lines.  They
appear to form two groups (Fig.~\ref{fg:asymm}).  \ha\ and \hb\ show
little or no asymmetry, while \civ\ and \lya\ show significant
asymmetry with excess flux in the blue wing.    These agree with the
results of \civ\ and \hb\ in \citet{BasLao05}, but we also show that
\civ\ and \lya\ asymmetries are marginally correlated ($r=0.53,
p=0.01$, Table~\ref{tb:corrasymm}), and the asymmetry parameters for
\ha\ and \hb\ show little correlation ($r=0.43, p=0.05$).

\subsection{Discussion of Line Profiles \label{sec:profile}}

The difference between high and low-ionization line profiles has been
investigated extensively in early studies
\citep[e.g.,][]{Gask82,Wilk84,Espe89,Corb91,TytFan92,Laor95,Sule95,Will95,CorBor96,Marz96,Vand01,Rich02}.
It is well known that, compared with low-ionization lines,
high-ionization lines tend to have large blue shifts and asymmetries
with stronger blue wings.  Our data give us the advantage of comparing, for
each QSO, all the strong UV and optical emission lines of different ionization
stage at essentially the same epoch.

In our sample, the decreasing significance of FWHM correlations with
\hb\ FWHM from low-ionization lines to high-ionization lines suggests
that, within the broad line emitting region, kinematics is a function
of ionization.  This is also supported by the rough sequence of line
peak shifts and asymmetries with ionization stage.  All the evidence
show two distinct groups of lines and a possible third group in
between.  High-ionization lines \civ\ and \lya\ are clearly distinct
from low-ionization \hb\ and \ha, while \mgii\ and \ciii\ seems
intermediate.

Systematically asymmetric line profiles and shifts must be the result
of radial motions, together with obscuration (optical depth, dust)
\citep[e.g.][]{Fer79}.  For example, stronger blue wings and
blueshifts could be the result of nuclear outflow, with emission from
the far side of the center suppressed.  Stronger blue wings could also
be the result of flow towards the nucleus with anisotropic emission
stronger from the more highly ionized regions facing the continuum
source.  
%Radial motion plus obscuration or anisotropic optical depth effects
%would predict that narrower lines, larger blue shifts and stronger
%blue wings go together.  
\citet[][]{Rich02} and \citet{Ric06} interpreted the range of CIV
profiles (blueshifts) in a large SDSS sample as a combination of dust
obscuration and orientation. 

While it seems that the kinematics of the BLR may be related to the
ionization structure, our sample shows that the high ionization \civ\
line is broader than the low ionization \hb\ line in only about half
the objects.  This appears to contradict a simple radially stratified
ionization structure of the BLR indicated by reverberation mapping, in
which high-ionization clouds are closer to the ionizing source and
therefore have higher (virial) velocity dispersion
\citep[e.g.,][]{Pete91,Kori95}.  However, the measured line widths may
not simply be a function of Keplerian velocity, and they can include
additional velocity components contributed by wind, outflows etc.
We previously noted that narrow line Seyfert 1 (NLS1) objects with
\hb\ FWHM  $< 2000$\,\kms showed \lya\ and \civ\ FWHM $>$\,\hb\ FWHM,
and argued for the presence of a high-ionization outflow in NLS1s
\citep{Wil00}.  \citet{BasLao05} have analyzed a larger sample and
shown that \hb\ is broader than \civ\ when \hb\ FWHM $>$ 4000 \kms.
They attribute this partly to an outflowing wind component in the BLR,
as we suggested for our sample \citep[][]{Wil00}.  \citet{VesPet06}
have re-analyzed the Baskin \& Laor sample, culling the less-reliable
data, and agree that there is probably a strong outflowing wind in
NLS1s \citep[see also][]{Ves04}.  Detailed studies of individual
NLSy1s have shown that high-ionization lines have a clear blueshifted
wind component, indicated by the large flux excess in their blue wing
\citep{LeiMoo04,Leig04, Yuan07}.

\hb\ FWHM and \ha\ FWHM have been used as a measure of BLR velocity
dispersion to estimate the black hole mass
\citep[e.g.,][]{Pete93,Kasp00,GreHo05}.  \mgii\ has also been used for
estimating black hole mass at higher redshifts \citep{McLJar02}, and
it seems to be valid as the FWHMs of \hb\ and \mgii\ are strongly
correlated.  However, although it also seems to work statistically,
the use of \civ\ \citep[e.g.,][]{Vest02} may introduce significant
uncertainty.  This is suggested by the evidence for outflow in NLS1s,
and is the reason \citet{VesPet06} exclude NLS1s from their
investigation of the use of \civ\ FWHM to estimate black hole mass.
Asymmetry and line shifts occur in \civ, other high-ionization lines
(\S\ref{sec:shift}), and to a lesser extent, in \hb, and not just for
NLS1s.  This suggests that further refinement in black hole mass
determinations may be possible after accounting for obscuration and
optical depth effects (by measuring the width of the unsuppressed wing
about the expected systemic velocity), or taking into account
additional non-virial (outflow) motions (measuring the narrower virial
wing about the expected systemic velocity).  \citet{VesPet06} note
that, statistically, the uncertainties in the above effects for
non-NLS1s are within the uncertainties of single-epoch width
measurements, thus validating the use of \civ\ for black hole mass
measurements.  However, for objects with an extreme outflow component
in the line profile (e.g., the aforementioned NLS1s), \civ\ still
could not be used in the black hole mass measurements.
%However, for some extreme objects, emission from outflows could well
%be important.

%%%%%%%%%%%%%%%%%%%%%%%%%%%%%%%%%%%

\section{SUMMARY\label{sec:summ}}

\begin{enumerate}

\item We present  quasi-simultaneous UV and optical spectra covering a
broad wavelength range from below \lya\ to at least \ha\ for an
essentially complete sample of low-redshift quasars.  We measured the
UV-optical continuum slopes (power-law).  Line widths, shifts, and
asymmetries are also measured for all strong emission lines and
results presented.

%\item Direct measurements and analyses of spectral properties confirm
%most results from the spectral principal component analysis.

\item Our analyses of UV-optical emission line profiles indicate
radial motions and anisotropic line emission (wind, optical depth
effects or dust obscuration) that is related to the ionization
structure, thus excluding a simple radially stratified ionization
structure, with gas in virial motion.

We will present detailed correlation analyses and multi-variate
analyses of all the spectral parameters in a separate paper
\citep{Will07}.

%\item We do not see obvious intrinsic reddening for the sample.

\end{enumerate}

\acknowledgments

Z. S. thanks Edward L. Robinson at the University of Texas at Austin
for his help on this project.  We thank the staffs of McDonald
Observatory, especially D. R. Doss, and of the Space Telescope Science
Institute for their expert guidance.  This work has been supported by
NASA under Grant No.  NNG05GD03G issued through the Office of Space
Science.  B. J. W.  acknowledges financial support by NASA through
LTSA grant NAG5-3431 and grant GO-06781 from the Space Telescope
Science Institute, which is operated by the Association of
Universities for Research in Astronomy, Inc., under NASA contract
NAS5-26555.  We are also grateful for support from the US National
Science Foundation, through Grant No.  AST-0206261 (B. J. W.) and
AST-0507781 (MSB), and from the National Natural Science Foundation
of China through Grant No. 10643001 (Z. S.).  This research has made
use of the NASA/IPAC Extragalactic Database (NED) which is operated by
the Jet Propulsion Laboratory, California Institute of Technology,
under contract with the National Aeronautics and Space Administration.

%and \hst\ grant HST-AR-09913.01-A.


%% copy editing.  Individual instruments can be provided in parentheses,
%% after the keyword, but they will not be verified.

Facilities: \facility{HST(FOS)}, \facility{McDonald}.

\appendix


\section{Notes on Data for Individual Objects \label{ap:data}}

\begin{figure}
%\plotone{wavecover.eps}
%\plotone{f1.eps}
\epsscale{.8}
\plotone{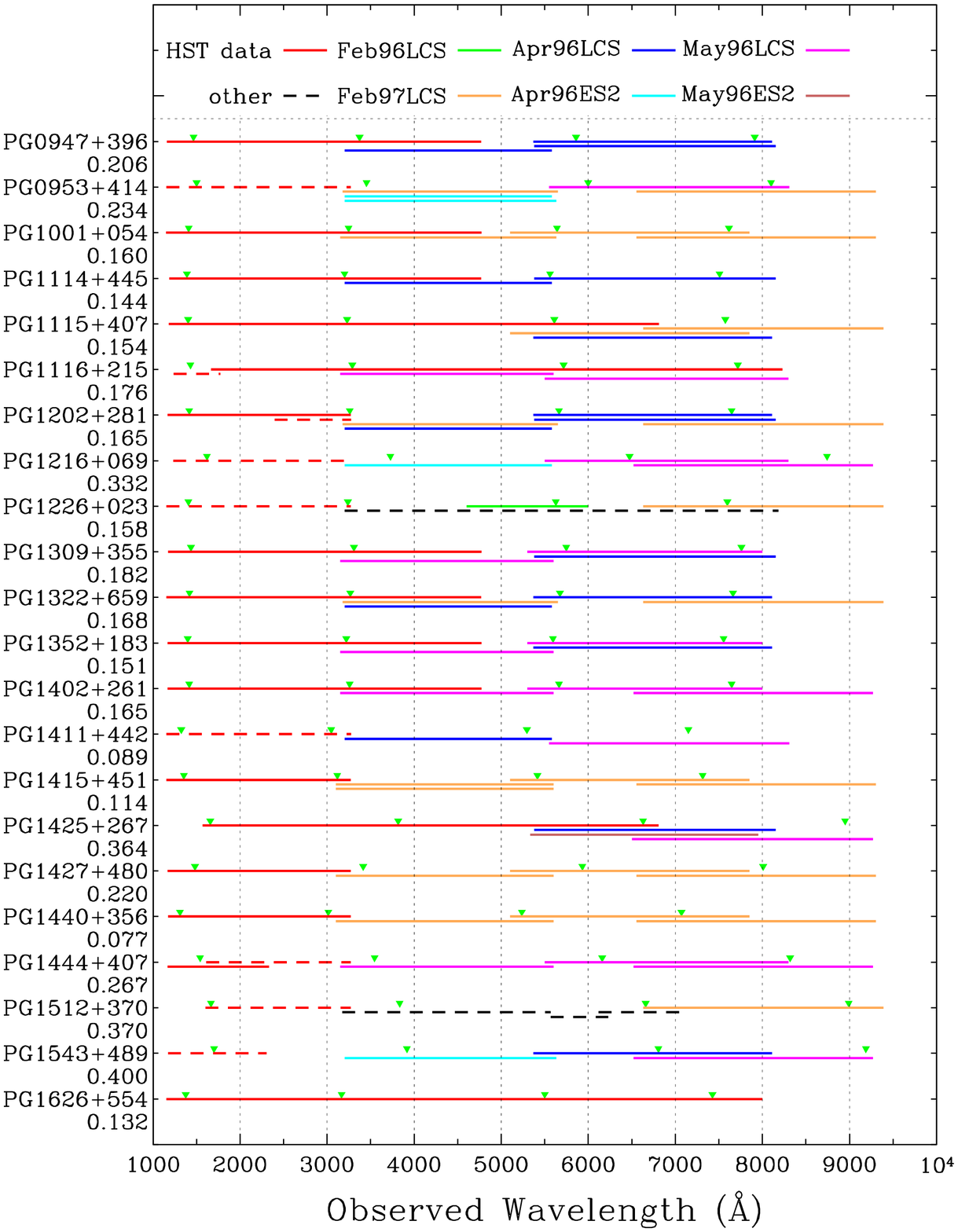}
\epsscale{1}
\caption{Summary of observations and wavelength coverage.  Object
names and redshifts are labeled on the left.  Each horizontal 
line indicates one individual spectrum; dashed lines
indicate archival data.  Different colors indicate different 
observing runs and instruments, Large Cassegrain Spectrograph 
(LCS) or Electronic Spectrograph 2 (ES2).
The small triangles mark, from left to
right, the locations of \lya, \mgii, \hb, and \ha. Some \hst\ spectra
extend to the optical.  Duplicate spectra with the same wavelength
coverage are all used in the final combination for
each object (\S\ref{sec:red:com}).
}
\label{fg:wavecover}
\end{figure}

The UV spectrum of each object is either from $HST$ archives or from
our own observations, and in general the optical spectra were obtained
at McDonald Observatory.  There are a few cases where we use data from
other sources (see also Figure~\ref{fg:wavecover} and
Table~\ref{tb:log}).  Wavelengths here are all in the observed frame.

\begin{description}

\item[PG1048+342] Due to the very low flux measured from IUE data,
this object would have required an unreasonable amount of \hst\
time to observe, and therefore the observation was not proposed.
It turned out later that the IUE data spectrum was weak, probably
because of a pointing problem.  This is the object that is in
the complete sample of \citet{L94,L97}, but is not included in
the analyses in this paper.

\item[PG1116+215] Most data (1668--8231\AA) are from our new
observations (both $HST$ and McDonald), while a small part of the
spectrum (1239--1774\AA, including \lya) is from the $HST$ archive.

\item[PG1202+281] Archival \hst\ UV data in the wavelength
range 2400--3277\AA\ are also used to
increase the signal-to-noise ratio.  The flux density of this spectrum
agrees very well with the new $HST$ spectrum, although this object
is highly variable \citep[][and references therein]{Sitko93}.

%%mostly continuum, only 2/3 of MgII line (blue) in common
%%Bev did this in /data/pan/bev/hst.ari/Images

\item[PG1226+023 (3C273)] Some optical data (3200--8183\AA) are from
1981 and 1988 observations using the UVITS spectrograph and image
dissector scanner (IDS) on the 2.7m telescope at McDonald Observatory
\citep*{Will85}.  

\item[PG1512+370] The blue part of our optical spectrum (shown in
Figure~\ref{fg:wavecover}, 3201--5631\AA) is not used due to poor
quality, instead, data are from observations by Jack Baldwin
(3174--5570\AA), Boroson and Green (BG92) (6120--7052\AA) and Bev
Wills' archival McDonald IDS data (5567--6230\AA).

\item[PG1543+489] We were not able to obtain the UV spectrum between
2307--3200\AA\ (G190H) because the data were supposed to be obtained
for another $HST$ proposal before ours, but were never obtained.  We
also do not have data for part of its \ha\ red wing
(Fig.~\ref{fg:spec}).

\end{description}

%%%%%%%%%%%%%%%%%%%%%%%%%%

%%%%%%%%%%%%%%%%%%%%%%%%%%%%%%%%%%%%%%%%%%%%%%%%%%%%%%%%%%%%
%%%%%%%%%%%%%%%%%%%%%%%%%%%%%%%%%%%%%%%%%%%%%%%%%%%%%%%%%%%%
%%%%%%%%%%%%%%%%%%%%%%%%%FFigures

%%%%%%%%%%%%%%%%%%%%%%%%%%%%%%%%%%%%%%%%%%% Cont
\begin{figure}
%\plotone{fitcont1.eps}
\epsscale{.8}
\plotone{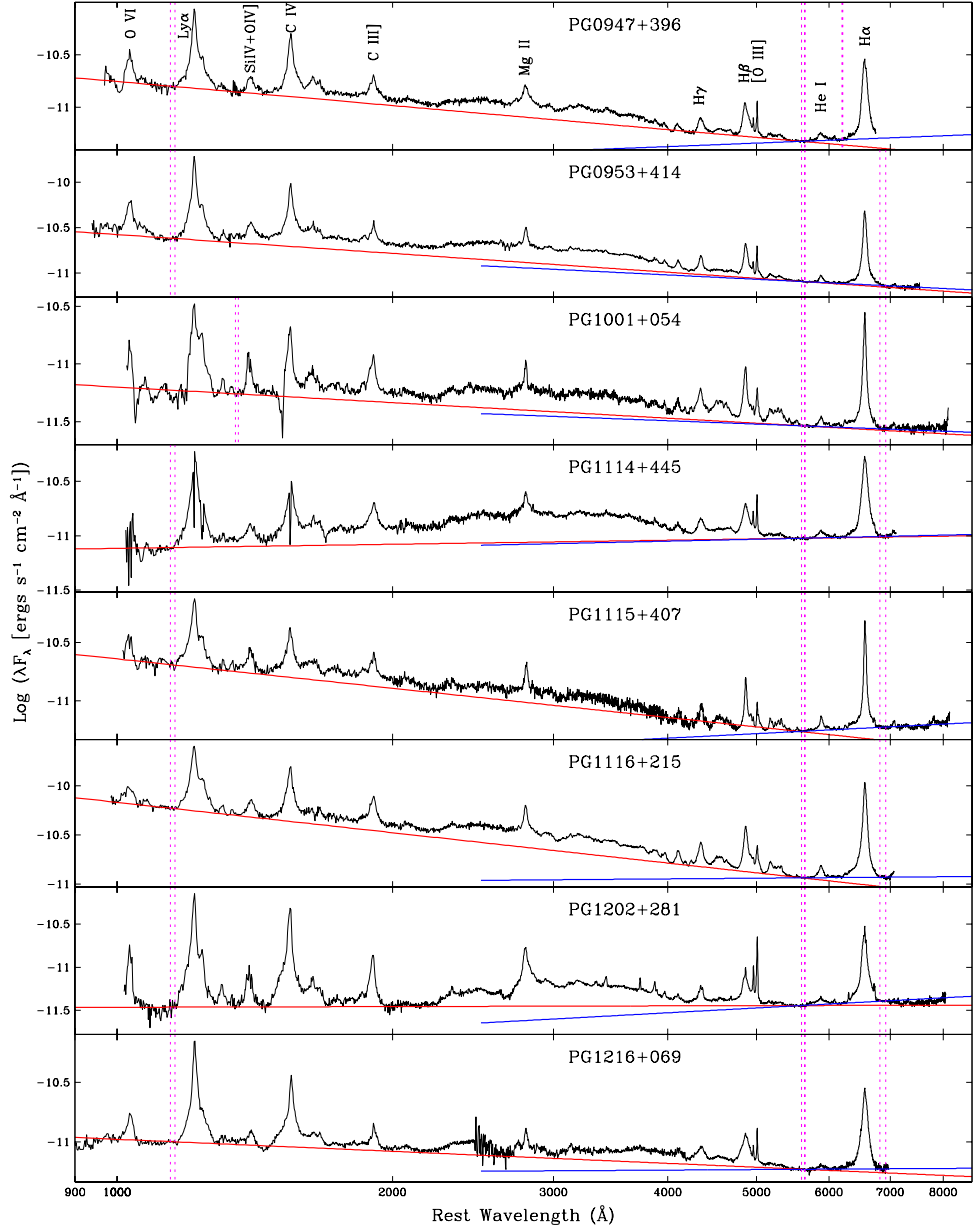}
\epsscale{1}
\caption{UV-optical spectra and fitted power-laws for different
regions.  The pairs of vertical dotted lines indicate
the continuum windows used for fitting
\auvoa\ ($\sim$1200--5500\AA) and
\auvob\ ($\sim$5500--8000\AA) (Table~\ref{tb:index}).
The \hst\ spectra are oversampled and 
have been rebinned to a resolution of 2\AA\
for display purpose.  The flux is still in observed frame.
[{\em See the electronic edition of the Journal for a color version of
this figure.}]
}
\label{fg:spec}
\end{figure}

\addtocounter{figure}{-1}
\begin{figure}
%\plotone{fitcont2.eps}
\plotone{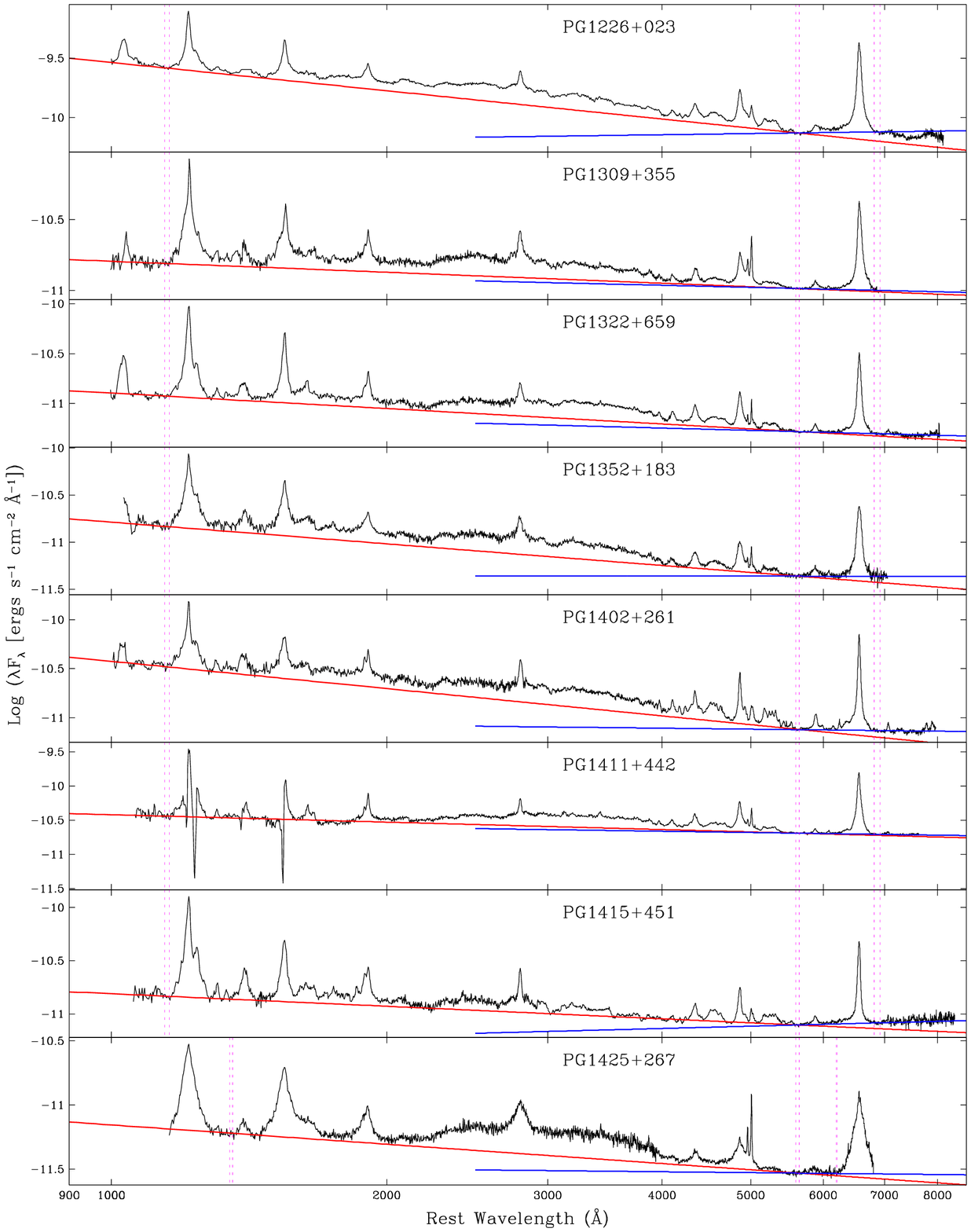}
\caption{\em Continued 
\label{fg:spec2}}
\end{figure}
 
\addtocounter{figure}{-1}
\begin{figure}
%\plotone{fitcont3.eps}
\plotone{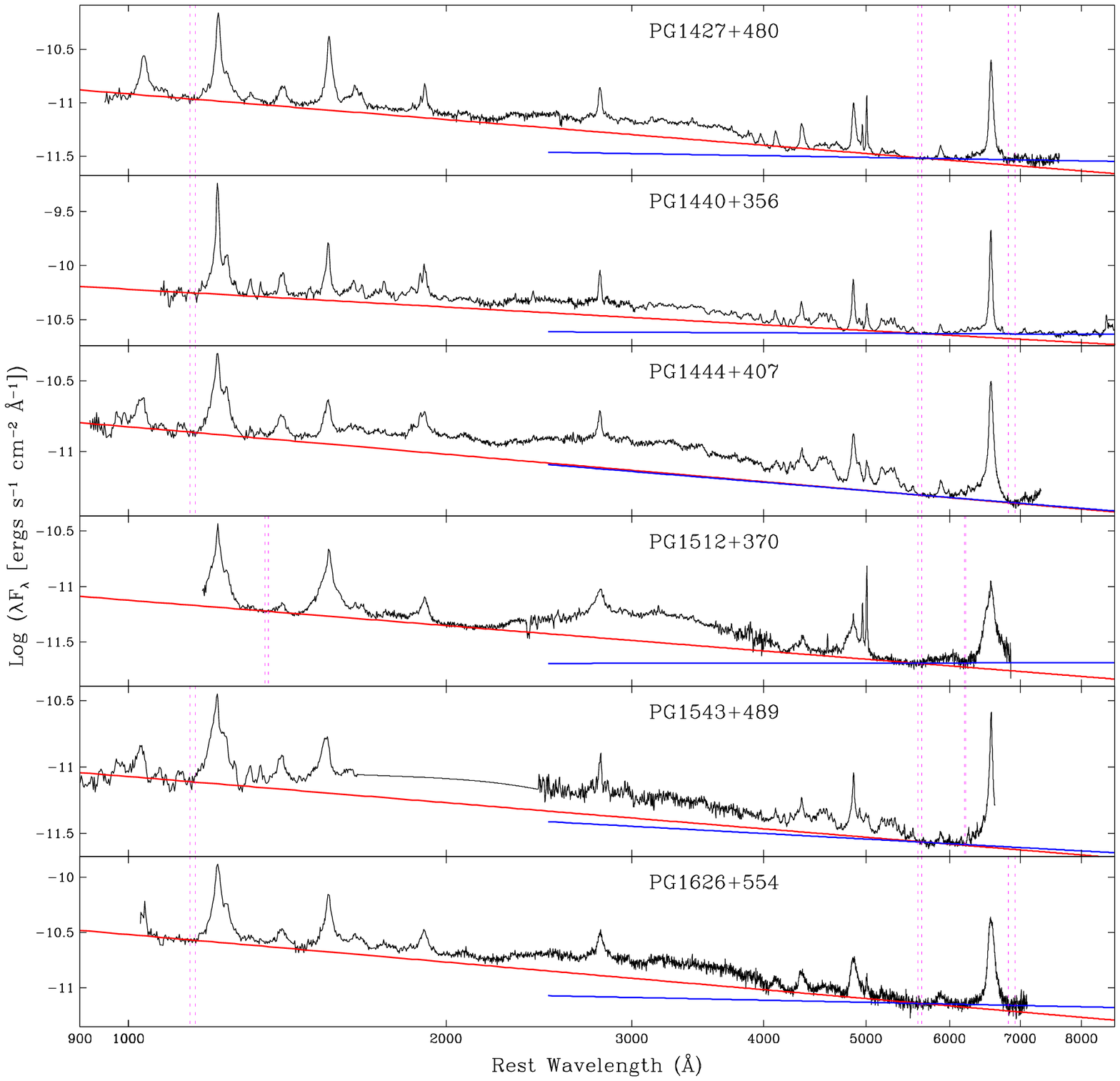}
\caption{\em Continued
\label{fg:spec3}}
\end{figure}

%%%%%%%%%%%%%%%%%%%%%%%%%%%%%%%%%%%%%%%%%%% fit line
\begin{figure}
%\plotone{fitline_color.eps}
\epsscale{.8}
\plotone{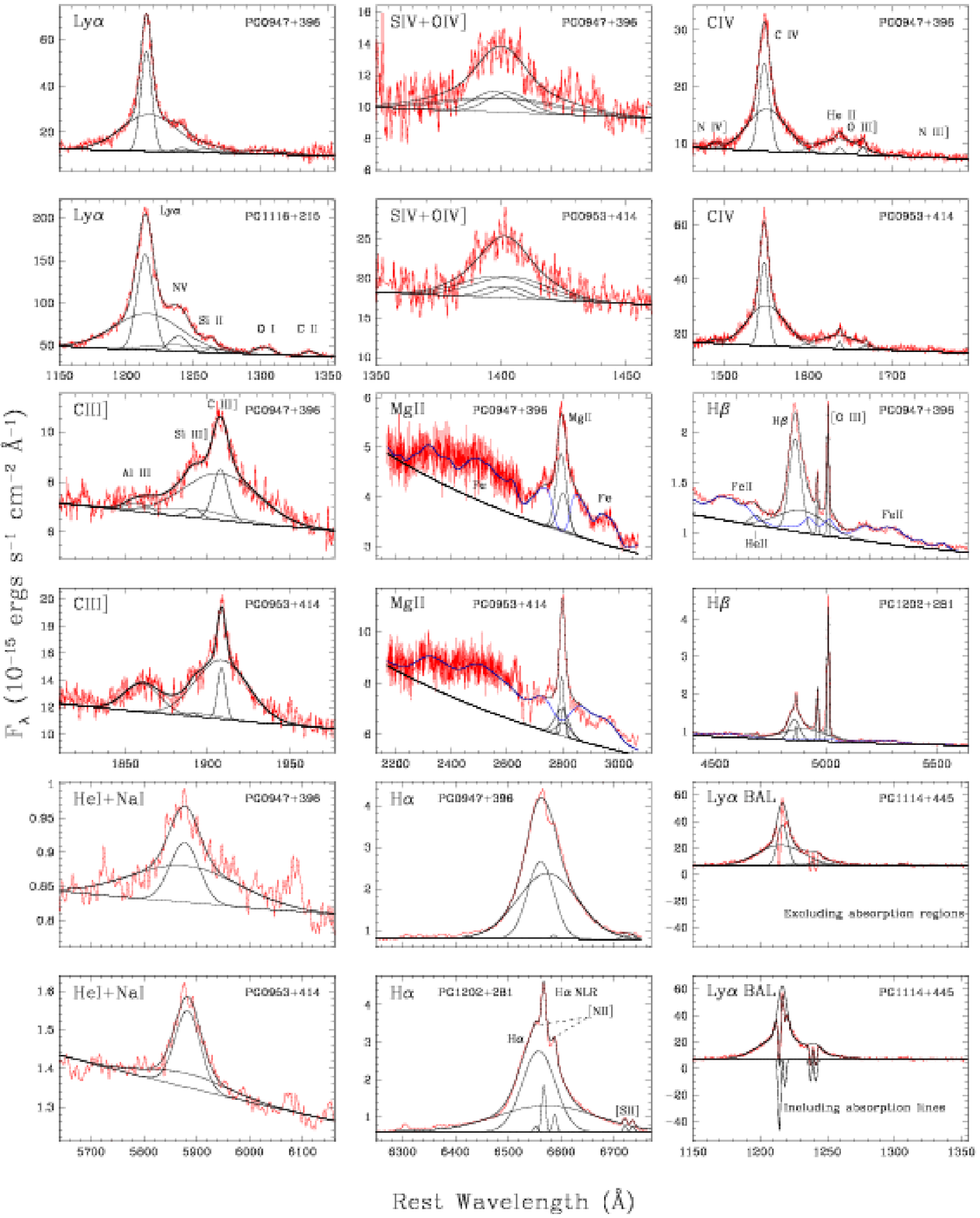}
%\plotone{f3.eps}
\epsscale{1}
\caption{Model fitting to the emission lines.  Two objects are shown
for each line region.  
Shown with the data are the fitting results (thick solid lines),
the local continua and individual components of the model (thin
lines).
The last two panels (lower-right corner) show \lya\ region of
PG~1114+445, where the BAL features are excluded (upper) and modeled
(lower) in the fitting, respectively.
[{\em See the electronic edition of the Journal for a color version of
this figure and for plots of the complete sample
(Figures~\ref{fg:fitline}.1--\ref{fg:fitline}.16)}]
}
\label{fg:fitline}
\end{figure}

%%%%%%%%%%%%%%%%%%%%%%%%%%%%%%%%%%%%%%%%%%% fit sbb
\begin{figure}
%\plotone{fitsbb.eps}
\plotone{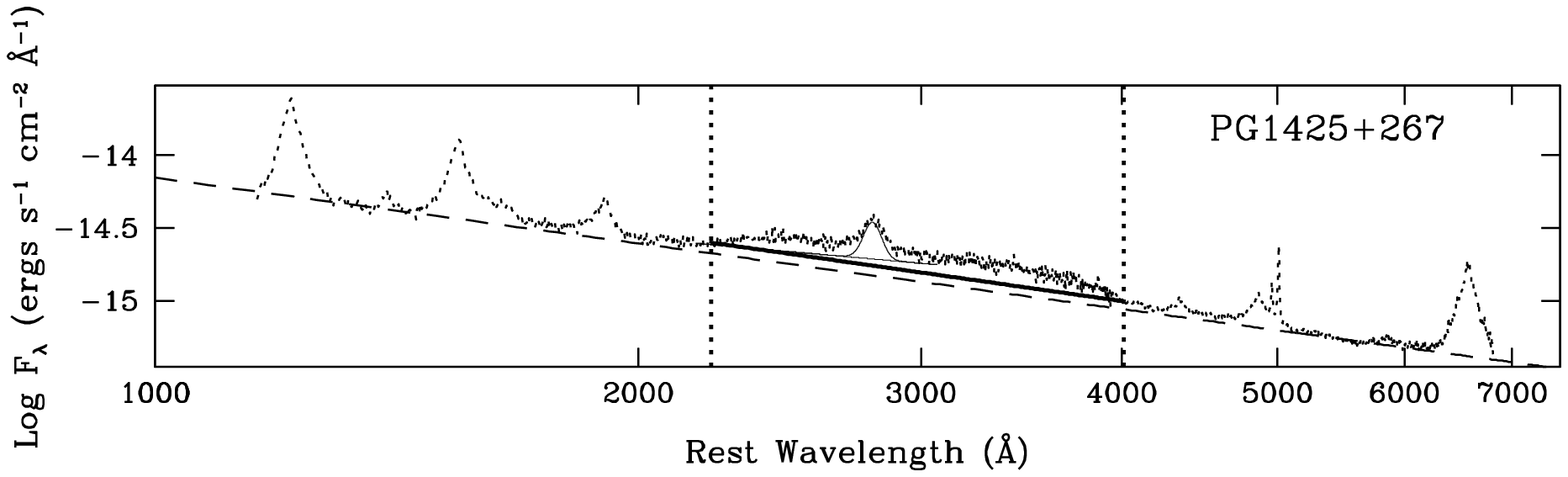}
\caption{Illustration of different continua defined for estimating the
flux of the Small Blue Bump (\S\ref{sec:measure:sbb}).  The global
power-law (\auvo, dashed-line) is shown as well as the local power-law
continuum (solid thick line) defined between 2220 and 4010\,\AA.  The
vertical lines indicate the flux integration region for the Small Blue
Bump and \mgii\ has been modeled and subtracted.  \mgii\ model and its
fitting continuum are overplotted (solid thin line).  
Note that \mgii\ fitting continuum
is not good to estimate the flux of the Small Blue Bump
(\S\ref{sec:measure:sbb}), 
}
\label{fg:fitsbb}
\end{figure}
%%%%%%%%%%%%%%%%% FWHM
\begin{figure}
\epsscale{0.7}
%\plotone{fwhm.eps}
\plotone{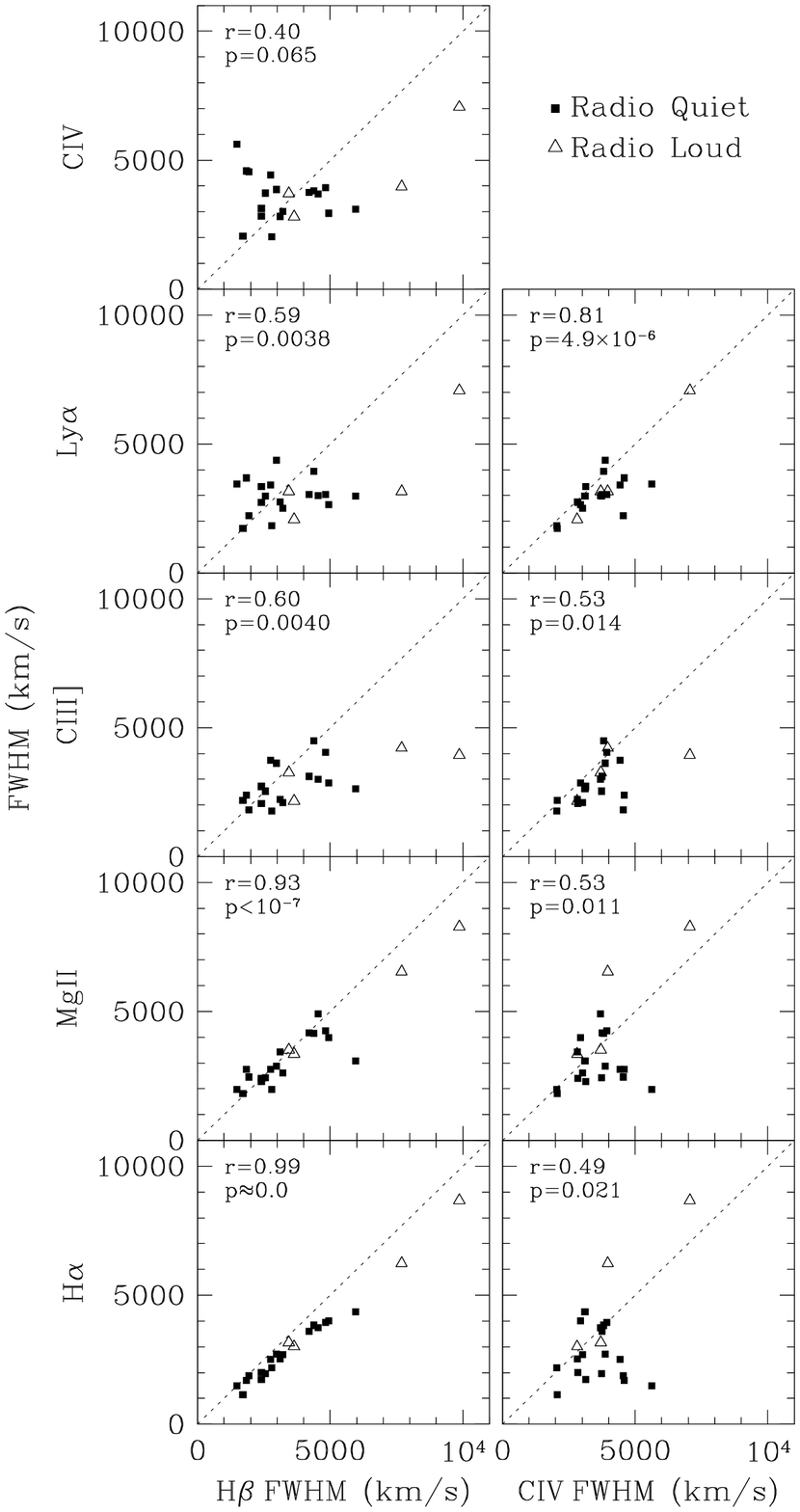}
\caption{Emission line FWHM vs. those of \hb\ and \civ.  The dotted
lines indicate equal FWHMs.  Note that
we do not have \ciii\ measurement for PG~1543+489.
}
\label{fg:fwhm}
\end{figure}

%\clearpage
%%%%%%%%%%%%%%%%% Shift

\begin{figure}
\epsscale{0.5}
%\plotone{shifthist.eps}
\plotone{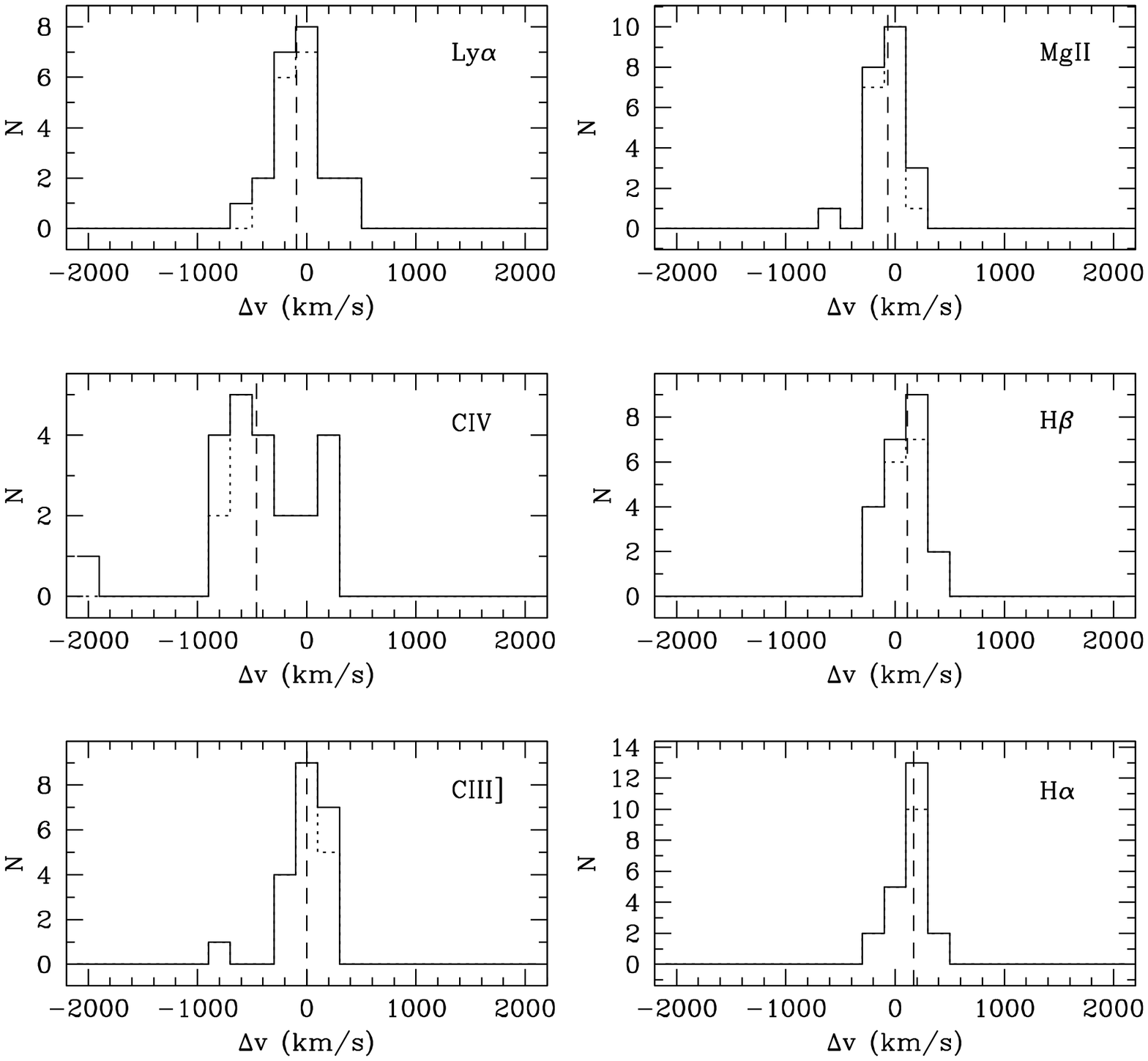}
\caption{Distribution of emission-line shift for the
whole sample (solid line) and for leaving out 3 objects (dotted line) 
whose redshifts are not measured using \oiii.
Positive values indicate redshift.
The vertical dashed lines indicate the median values for the whole
sample.
}
\label{fg:shift}
\end{figure}
\notetoeditor{One column please}

%\clearpage
%%%%%%%%%%%%%%%%% Asymmetry
\begin{figure}
\epsscale{0.5}
%\plotone{asymmhist.eps}
\plotone{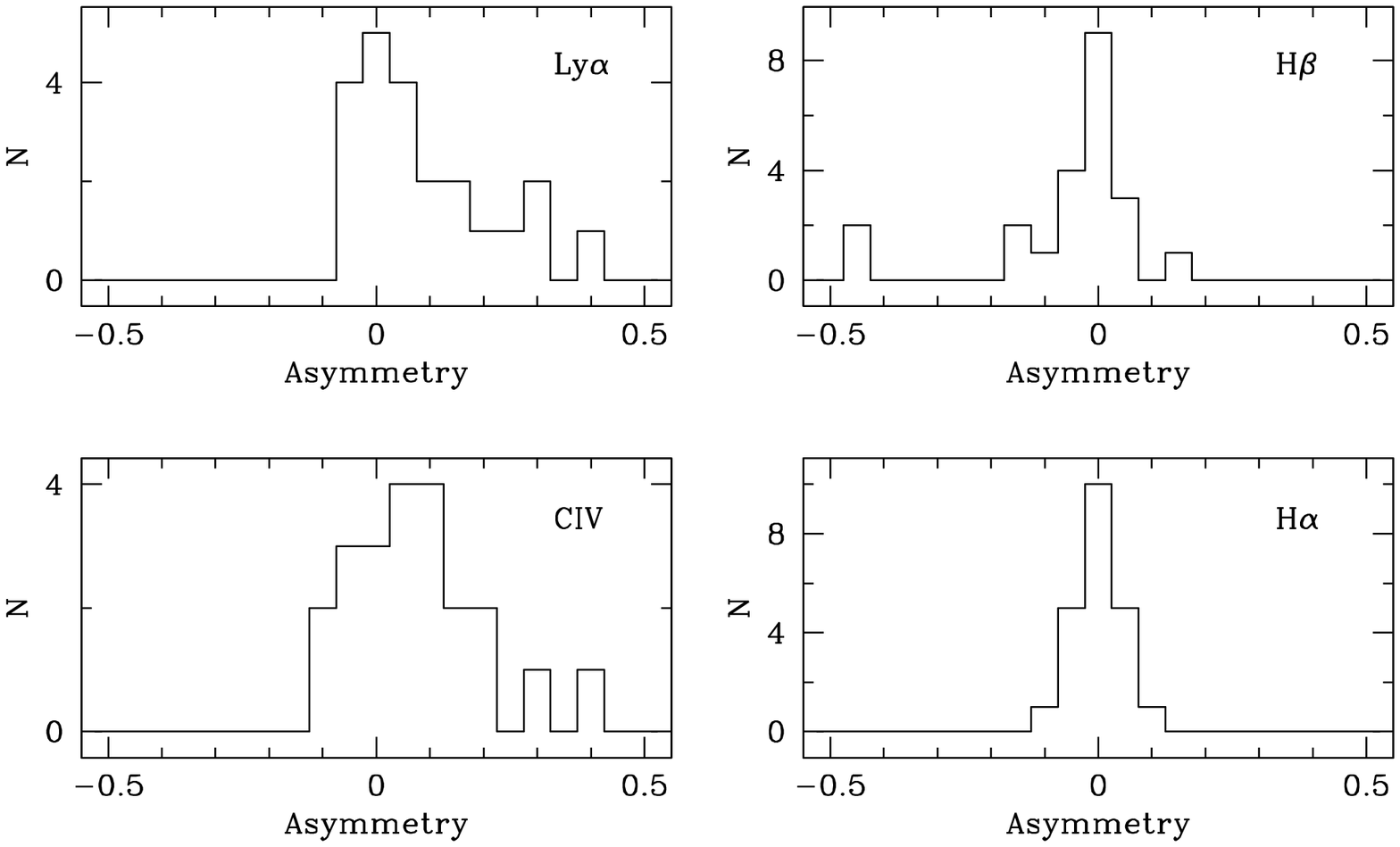}
\caption{Distribution of emission-line asymmetry.
See \S\ref{sec:measure:line} for definition of the asymmetry parameter.
Positive values indicate excess flux in the blue wing.
\label{fg:asymm}
}
\end{figure}
%%%%%%%%%%%%%%%%%%%%%%%%%%%%%%%%%%%%%%%%%%%%%%%%%%%%%%%%%%%%
%%%%%%%%%%%%%%%%%%%%%%%%%%%%%%%%%%%%%%%%%%%%%%%%%%%%%%%%%%%%
%%%%%%%%%%%%%%%%%%%%%TTables

%%%%
\begin{deluxetable}{lllrlrc}
\tabletypesize{\footnotesize}
%\rotate
\tablecolumns{7}
\tablewidth{0pc}
\tablecaption{The Sample \label{tb:sample}}
\tablehead{
\colhead{Object} & \colhead{Other Name}  &
\colhead{$z$\tablenotemark{a}}
& \colhead{$m_B$\tablenotemark{b}}
& \colhead{\alphax\tablenotemark{c}} &
\colhead{$R$\tablenotemark{c}}
& \colhead{E(B-V)\tablenotemark{d}}
%& \colhead{($10^{20}{\rm cm}^{-2}$)}
}
\startdata
PG0947+396	&	        &	0.2056	&	16.40
&	$-$1.510	&	0.25	&	0.019	\\
PG0953+414	&	    	&	0.2341	&	15.05
&	$-$1.570	&	0.44	&	0.013	\\
PG1001+054	&		&	0.1603\tablenotemark{e}	&	16.13	&
$-$2.800	&	0.5	&	0.016	\\
PG1114+445	&		&	0.1440	&	16.05	&
$-$1.550	&	0.13	&	0.016	\\
PG1115+407	&		&	0.1541	&	16.02	&
$-$1.890	&	0.17	&	0.016	\\
PG1116+215	&	   TON 1388	&	0.1759	&	15.17
&	$-$1.730	&	0.72	&	0.023	\\
PG1202+281	&	    GQ COM	&	0.1651	&	15.02
&	$-$1.220	&	0.19	&	0.021	\\
PG1216+069	&		&	0.3319	&	15.68	&
$-$1.360	&	1.65	&	0.022	\\
\enddata
\tablenotetext{a}{From measurements of \oiii\ after removing
\feii\ emission in our separate higher resolution spectra, unless
noted (\S\ref{sec:redshift})}

\tablenotetext{b}{$B$ magnitude, from \citet{SchGre83}}

\tablenotetext{c}{From \citet[ and references therein]{L97}.
\alphax\ is soft X-ray spectral index (0.2--2 \kev), $f_\nu \propto \nu^{\alpha_X}$;
$R$ is radio (6 cm) to
optical (4400\AA) flux ratio in rest-frame.}

\tablenotetext{d}{From NED \url{(http://nedwww.ipac.caltech.edu/)} based on \citet{Schl98}.}
\tablenotetext{e}{From measurements of \oiii\ at the typical
resolution of spectra in this study 
(\S\ref{sec:redshift}).}

\tablenotetext{f}{From measurements of \hb\ and other emission lines
in the spectra of this study.  These objects have very weak \oiii. }

\end{deluxetable}

%%%%%%%%%%%%%%%%% obs log

\begin{deluxetable}{rrrrr@{}p{0pt}@{}rrrr@{}r@{}rrrr}
\tabletypesize{\scriptsize}
\tablecolumns{15}
\tablewidth{0pc}
\tablecaption{Observing Time, Observing Time Gap, and Scaling Factor
\label{tb:log}}
\tablehead{
\colhead{} & 
\multicolumn{4}{c}{Observing Date (UT)} & \colhead{} &
\multicolumn{4}{c}{Observing Time Gap (days)\tablenotemark{a}} & \colhead{} &
\multicolumn{4}{c}{Scaling Factor}
\\ \cline{2-5} \cline{7-10} \cline{12-15}
\colhead{} & \colhead{} &
\multicolumn{3}{c}{Ground-based} &
\colhead{} & \colhead{} &
\multicolumn{3}{c}{Ground-based} &
\colhead{} & \colhead{} &
\multicolumn{3}{c}{Ground-based} 
\\\cline{3-5} \cline{8-10}  \cline{13-15}
\colhead{Object} &
\colhead{\hst} & \colhead{Blue} & \colhead{Middle} & \colhead{Red} &
\colhead{} &
\colhead{\hst} & \colhead{Blue} & \colhead{Middle} & \colhead{Red} &
\colhead{} &
\colhead{\hst} & \colhead{Blue} & \colhead{Middle} & \colhead{Red}
}
\startdata
PG0947+396	&	1996-05-06	&	1996-04-19	&
1996-04-17	&		&	&	19	&	2
&	0	&		&	&	0.97	&	0.99
&	1	&		\\
	&		&		&	1996-04-18	&
&	&		&		&	1	&
&	&		&		&	1.04	&
\\
PG0953+414	&	1991-06-18	&	1996-04-24	&
1996-05-13	&	1997-02-11	&	&	-1791	&
-19	&	0	&	274	&	&	1.29	&
0.94	&	1	&	1.08	\\
	&	1991-11-05	&	1996-04-25	&
&		&	&	-1651	&	-18	&
&		&	&	1.29	&	1.22	&
&		\\
	&		&	1997-02-10	&		&
&	&		&	273	&		&
&	&		&	0.81	&		&
\\
PG1001+054	&	1997-01-15	&	1997-02-10	&
1997-02-13	&	1997-02-16	&	&	-26	&
0	&	3	&	6	&	&	0.93	&
1	&	1.05	&	0.97	\\
PG1114+445	&	1996-05-13	&	1996-04-19	&
1996-04-17	&		&	&	24	&	0
&	-2	&		&	&	1.08	&	1
&	1.03	&		\\
PG1115+407	&	1996-05-19	&		&
1996-04-18	&	1997-02-16	&	&	31	&
&	0	&	304	&	&	1.16	&
&	1	&	0.97	\\
\enddata
\tablecomments{See
Figure~\ref{fg:wavecover} for wavelength coverage for each observation.}
\tablenotetext{a}{Relative to the reference spectrum which has a time gap of
zero and a scaling factor of 1.}
\tablenotetext{b}{There may be large uncertainty in these old 
archival data.  The flux calibration was not
good and the applied scaling factors can be misleading.
}
\end{deluxetable}

%\clearpage
%%%%%%%%%%%%%%%%%%%%%%%%%%%%%%%%%%%%%%%%%%%%%%%% continuum parameters
\begin{deluxetable}{rcrrrclrrrrr}
%\tabletypesize{\tiny}
\tabletypesize{\scriptsize}
%\tabletypesize{\small}
\tablecolumns{12}
\tablewidth{0pc}
\tablecaption{Spectral Indices \label{tb:index}
\label{tb:cont}}
\tablehead{
&\multicolumn{3}{c}{1200--5500\AA\ region}
&& \multicolumn{3}{c}{$\sim$5500--8000\AA\ region}& \\ \cline{2-4} \cline{6-8}
\colhead{Object}&      
%\colhead{CWindows\tablenotemark{a}}&      
\colhead{Win\tablenotemark{a}}&      
\colhead{$f_{1000}$\tablenotemark{b}} &
\colhead{\auvo}& 
&
%\colhead{CWindows\tablenotemark{a}}&      
\colhead{Win\tablenotemark{a}}&      
\colhead{$f_{1000}$\tablenotemark{b}} &
\colhead{\aored}& 
\colhead{$\Delta \alpha$\tablenotemark{c}}&
\colhead{$f_\nu$(1keV)\tablenotemark{d}}& 
\colhead{\axnu\tablenotemark{d}}&
\colhead{\aox\tablenotemark{e}}
}
\startdata
PG0947& a,d& $  17.6\pm   0.9$& $ -1.76\pm  0.04$ &
	&d,e& $\phn   2.6^{+2.0  }_{-1.1  }$ 
	& $ -0.66 \pm 0.32 $ & $ -1.10$
	&  7.64& $ -1.51$& $ -1.42$ \\
PG0953& a,d& $  26.5\pm   1.1$& $ -1.69\pm  0.03$ &
	&d,f& $  18.9^{+6.4  }_{-4.8  }$ 
	& $ -1.50 \pm 0.16 $ & $ -0.19$
	&  9.33& $ -1.57$& $ -1.47$ \\
PG1001& b,d& $   6.3\pm   0.4$& $ -1.45\pm  0.06$ &
	&d,f& $\phn   4.9^{+6.3  }_{-2.7  }$ 
	& $ -1.30 \pm 0.45 $ & $ -0.15$
	&  0.05& $ -2.80$& $ -2.29$ \\
PG1114& a,d& $   7.7\pm   0.3$& $ -0.88\pm  0.03$ &
	&d,f& $\phn   6.9^{+2.3  }_{-1.7  }$ 
	& $ -0.81 \pm 0.16 $ & $ -0.06$
	&  1.91& $ -1.55$& $ -1.65$ \\
PG1115& a,d& $  22.8\pm   1.2$& $ -1.83\pm  0.04$ &
	&d,f& $\phn   2.8^{+0.9  }_{-0.7  }$ 
	& $ -0.60 \pm 0.15 $ & $ -1.23$
	&  3.79& $ -1.89$& $ -1.61$ \\
PG1116& a,d& $  68.0\pm   2.0$& $ -2.03\pm  0.02$ &
	&d,f& $  10.2^{+2.9  }_{-2.2  }$ 
	& $ -0.93 \pm 0.14 $ & $ -1.10$
	& 13.10& $ -1.73$& $ -1.54$ \\
PG1202& a,d& $   3.4\pm   0.3$& $ -0.97\pm  0.07$ &
	&d,f& $\phn   1.3^{+0.5  }_{-0.4  }$ 
	& $ -0.42 \pm 0.18 $ & $ -0.55$
	&  9.88& $ -1.22$& $ -1.19$ \\
PG1216& a,d& $  10.5\pm   0.4$& $ -1.34\pm  0.03$ &
	&d,f& $\phn   5.4^{+2.5  }_{-1.7  }$ 
	& $ -0.95 \pm 0.21 $ & $ -0.38$
	&  7.93& $ -1.36$& $ -1.37$ \\
\enddata
\tablecomments{\auvo\ and \aored\ are rest-frame \alphal\ 
($f_\lambda \propto \lambda^{\alphal}$); 
\ax\ and \aox\ (between 2500\AA\ and 2\,keV) are rest-frame \anu
($f_\nu \propto \nu^{\anu}$). 
}
\tablenotetext{a}{Continuum windows used for fitting UV-optical
spectra 
(\S\ref{sec:measure:index}, Fig.~\ref{fg:spec}).
a:1144\AA--1157\AA;
b:1348\AA--1358\AA;
c:4200\AA--4230\AA;
d:5600\AA--5648\AA;
e:6198\AA--6215\AA;
f:6820\AA--6920\AA.
}
\tablenotetext{b}{$f_{1000}$ --- fitted continuum flux
density at 1000\AA\ for corresponding regions ($10^{-15}$ \fluxl).}
\tablenotetext{c}{$\Delta\alpha = \auvo - \aored$}
\tablenotetext{d}{From \citet{L97}. 
$f_\nu$(1keV) --- $f_\nu$ at 1\,keV 
(in units of $6.626\times10^{-31}$ \fluxhz).  }
\tablenotetext{e}{\aox\ -- \anu\ between 2\,keV and 2500\AA.}

\end{deluxetable}

%\clearpage
%%%%%%%%%%%%%%%%%%%%%%%%%%%%%%%%%%%%%%%%%%%%%%%% line fitting parameters
\begin{deluxetable}{rllccccr}
\tabletypesize{\tiny}
%\tabletypesize{\scriptsize}
%\tabletypesize{\small}
\tablecolumns{7}
\tablewidth{0pc}
%\tablecaption{Gaussian Components and Free Parameters for Fitting Emission
%Lines in Different Regions
\tablecaption{Fitting Parameters for Emission Lines
\label{tb:fitpar}}
\tablehead{
&&      &Gaussian& 
\multicolumn{3}{c}{Gaussian Parameter\tablenotemark{b}}& 
\colhead{Flux}\\ \cline{5-7}
\colhead{ID}&      
\colhead{Line}&      
\colhead{$\lambda$\tablenotemark{a}} &
\colhead{Component}& 
\colhead{Flux}&
\colhead{Center}&
\colhead{FWHM}&
\colhead{Ratio\tablenotemark{c}}
}
\startdata
%%%%%%	Lya
1 & \lya&1215.67&narrow&free&free&free&\\
2 & &&broad&free&free&free&\\
3 & NV&1238.81&narrow&free&1&1&1\\
4 & &&broad&\lya\ shape\tablenotemark{d}&2&2&1\\
5 & NV&1242.80&narrow&3&3&3&1\\
6 & &&broad&4&4&4&1\\
7 & SiII&1260.42&  &free&1&free&0.33\\
8 & SiII&1264.73&  &7&7&7&0.6\\
\enddata

\tablecomments{Each section separated by a horizontal line is for
one emission line region.  All parameters are related within the same region
only unless noted.}
%%%Bug in tablenotetext: Can't start with 'L'.
\tablenotetext{a}{.Laboratory wavelength.  In vacuum for 
$\lambda < 2000$\,\AA, in air for $\lambda > 2000$\,\AA.  The actual combined
spectra are in vacuum wavelength scale.}
\tablenotetext{b}{A number indicates, in the same section, 
the ID number of a component to which the parameter is tied.}
\tablenotetext{c}{.Line component intensity ratio within multiplets, from
\citet{Bald96}.  Narrow and broad components are not related.}
\tablenotetext{d}{The shape of the entire line (narrow+broad) is
tied to this line.}
\tablenotetext{e}{We fit this feature with \hei\ wavelength, but the
results should be considered as \nail+\heil\ since we cannot deblend
them.}
\tablenotetext{f}{The width is the same as \oiii.}

\end{deluxetable}

%%%%%%%%%%%%%%%%%%%%%%%%%%%%%%%%%%%%%%%%%%%%%%%%%%%%% Strong Line Measurements
% \newcounter{strongline}
% \addtocounter{strongline}{1}
% \renewcommand{\thetable}{\arabic{table}\Alph{strongline}}

%\setcounter{table}{\Alph{strongline}}
\begin{deluxetable}{clllrrp{0pt}@{}llp{0pt}@{}lllr}
\tabletypesize{\scriptsize}
\rotate
\tablecolumns{17}
\tablewidth{0pc}
\tablecaption{Parameters of Strong Emission Lines \lya, \nv, and
\siivoivl.
\label{tb:stronglinesA}}
\tablehead{
\colhead{} & 
\multicolumn{5}{c}{\lyal} & \colhead{} &
\multicolumn{2}{c}{\nvl\tablenotemark{a}} & \colhead{} &
\multicolumn{4}{c}{\siivl\ or \oivl\tablenotemark{b}}
\\ \cline{2-6} \cline{8-9} \cline{11-14}
\colhead{Object} &
\colhead{Flux}&\colhead{EW}&\colhead{FWHM}&\colhead{\dv}&\colhead{Asymm} &
\colhead{} &
\colhead{Flux}&\colhead{EW}&
\colhead{} &
\colhead{Flux}&\colhead{EW}&\colhead{FWHM}&\colhead{\dv}
}
\startdata
PG0947	&    1275$^{+75   }_{-64   }$
	&   107.5$^{+11.6}_{-10.6}$
	&   3000$^{+25     }_{-25     }$
	&$   -165$
	&$ -0.068^{+0.010  }_{-0.014  }$
&
	&\phn      68$^{+92   }_{-43   }$
	&\phn     6.0$^{+8.7 }_{-8.0 }$
&
	&\phn      75$^{+48   }_{-31   }$
	&\phn     7.7$^{+5.6 }_{-5.1 }$
	&   5560$^{+1170   }_{-920    }$
	&$    105$
\\
PG0953	&    2394$^{+79   }_{-69   }$
	&   125.5$^{+9.2 }_{-8.6 }$
	&   2500$^{+15     }_{-10     }$
	&$   -215$
	&$  0.014^{+0.003  }_{-0.005  }$
&
	&     310$^{+126  }_{-91   }$
	&    16.5$^{+7.7 }_{-7.1 }$
&
	&     121$^{+52   }_{-38   }$
	&\phn     6.9$^{+3.4 }_{-3.1 }$
	&   5430$^{+845    }_{-750    }$
	&$    420$
\\
PG1001	&\phn     432$^{+17   }_{-14   }$
	&   107.7$^{+10.6}_{-9.5 }$
	&   3350$^{+55     }_{-50     }$
	&$   -465$
	&$  0.049^{+0.003  }_{-0.002  }$
&
	&     215$^{+30   }_{-24   }$
	&    54.3$^{+10.9}_{-9.8 }$
&
	&\phn      52$^{+13   }_{-9    }$
	&    13.4$^{+4.1 }_{-3.8 }$
	&   3340$^{+155    }_{-155    }$
	&$  -1010$
\\
PG1114	&\phn     969$^{+36   }_{-31   }$
	&   146.7$^{+12.6}_{-11.5}$
	&   3040$^{+15     }_{-20     }$
	&$    250$
	&$  0.095^{+0.006  }_{-0.006  }$
&
	&     132$^{+56   }_{-40   }$
	&    20.1$^{+9.8 }_{-9.0 }$
&
	&\phn      55$^{+25   }_{-18   }$
	&\phn     8.7$^{+4.5 }_{-4.2 }$
	&   5815$^{+1165   }_{-920    }$
	&$    265$
\\
PG1115	&    1140$^{+109  }_{-91   }$
	&\phn    76.2$^{+11.8}_{-10.7}$
	&   3685$^{+60     }_{-65     }$
	&$   -185$
	&$  0.235^{+0.029  }_{-0.030  }$
&
	&     227$^{+175  }_{-108  }$
	&    15.7$^{+13.6}_{-12.2}$
&
	&\phn      91$^{+48   }_{-34   }$
	&\phn     7.1$^{+4.3 }_{-3.9 }$
	&   4665$^{+850    }_{-715    }$
	&$   -180$
\\
PG1116	&    4379$^{+171  }_{-147  }$
	&\phn    98.7$^{+6.2 }_{-5.9 }$
	&   4375$^{+20     }_{-30     }$
	&$   -385$
	&$ -0.035^{+0.002  }_{-0.002  }$
&
	&     777$^{+268  }_{-196  }$
	&    18.1$^{+6.8 }_{-6.5 }$
&
	&     262$^{+70   }_{-55   }$
	&\phn     7.5$^{+2.2 }_{-2.1 }$
	&   5265$^{+465    }_{-420    }$
	&$    380$
\\
\enddata
\tablecomments{Flux --- observed-frame flux in 10$^{-15}$\flux.
EW --- rest-frame equivalent width.
FWHM --- in \kms.
Asymm --- asymmetry parameter defined as 
$\mbox{Asymm} =
[{\lambda_c(\frac{3}{4}) - \lambda_c(\frac{1}{4})}] / {\mbox{FWHM}}
$ (\S\ref{sec:measure:line}).
\dv\ --- line peak velocity shift (\kms) relative to the systematic redshift
(\S\ref{sec:redshift}). A negative value indicates a blueshift of the
line peak.  The error for \dv\ depends on the uncertainty of the
redshift (\S\ref{sec:measure:error}).
}
\tablenotetext{a}{The values are for the sum of the doublet.  Each
single line is assumed to have the same shape as \lya\
(Table~\ref{tb:fitpar}).}
\tablenotetext{b}{\siivl\ and \oivl\ are assumed to have identical
profiles.  Values for PG1226 are measured from a separate spectrum
\citep{Shan05}.}
\end{deluxetable}

%%%%%%%%%%%%%%%%%%%	strong B
% \addtocounter{table}{-1}
% \addtocounter{strongline}{1}
\begin{deluxetable}{clllrrp{0pt}@{}lllrp{0pt}@{}ll}
\tabletypesize{\scriptsize}
\tablecolumns{16}
\tablewidth{0pc}
\tablecaption{Parameters of Strong Emission Lines 
\civ, \ciii, and \siiii
\label{tb:stronglinesB}}
\tablehead{
\colhead{} & 
\multicolumn{5}{c}{\civl\tablenotemark{a}} & \colhead{} &
\multicolumn{4}{c}{\ciiil} & \colhead{} &
\multicolumn{2}{c}{\siiiil\tablenotemark{b}}
\\ \cline{2-6} \cline{8-11} \cline{13-14}
\colhead{Object} &
\colhead{Flux}&\colhead{EW}&\colhead{FWHM}&\colhead{\dv}&\colhead{Asymm} &
\colhead{} &
\colhead{Flux}&\colhead{EW}&\colhead{FWHM}&\colhead{\dv}&
\colhead{} &
\colhead{Flux}&\colhead{EW}%&\colhead{FWHM}&\colhead{\dv}
}
\startdata
PG0947	&\phn     672$^{+36   }_{-30   }$
	&\phn    77.7$^{+5.5 }_{-5.2 }$
	&   3690$^{+75     }_{-70     }$
	&$   -265$
	&$ -0.089^{+0.014  }_{-0.014  }$
&
	&\phn     128$^{+20   }_{-16   }$
	&    19.8$^{+3.5 }_{-3.4 }$
	&   3000$^{+170    }_{-145    }$
	&$   -125$
&
	&\phn      23$^{+16   }_{-11   }$
	&\phn     3.5$^{+2.6 }_{-2.5 }$
\\
PG0953	&    1237$^{+61   }_{-51   }$
	&\phn    79.7$^{+5.2 }_{-4.9 }$
	&   3015$^{+55     }_{-50     }$
	&$   -310$
	&$ -0.120^{+0.017  }_{-0.011  }$
&
	&\phn     198$^{+22   }_{-19   }$
	&    17.8$^{+2.4 }_{-2.3 }$
	&   2105$^{+185    }_{-150    }$
	&$    -15$
&
	&\phn      12$^{+19   }_{-9    }$
	&\phn     1.1$^{+1.7 }_{-1.7 }$
\\
PG1001	&\phn     272$^{+20   }_{-16   }$
	&\phn    75.1$^{+7.3 }_{-6.8 }$
	&   3130$^{+80     }_{-80     }$
	&$   -695$
	&$  0.305^{+0.074  }_{-0.097  }$
&
	&\phn \phn      79$^{+9    }_{-8    }$
	&    26.9$^{+4.0 }_{-3.8 }$
	&   2725$^{+60     }_{-65     }$
	&$   -205$
&
	&\phn      32$^{+9    }_{-7    }$
	&    10.7$^{+3.5 }_{-3.3 }$
\\
PG1114	&\phn     498$^{+36   }_{-30   }$
	&\phn    81.1$^{+7.9 }_{-7.3 }$
	&   3935$^{+140    }_{-135    }$
	&$    -40$
	&$  0.053^{+0.018  }_{-0.018  }$
&
	&\phn     152$^{+17   }_{-14   }$
	&    25.3$^{+3.3 }_{-3.2 }$
	&   4045$^{+145    }_{-135    }$
	&$    155$
&
	&\phn      25$^{+14   }_{-10   }$
	&\phn     4.1$^{+2.4 }_{-2.4 }$
\\
PG1115	&\phn     519$^{+70   }_{-54   }$
	&\phn    47.6$^{+7.9 }_{-7.2 }$
	&   4585$^{+315    }_{-285    }$
	&$   -600$
	&$  0.095^{+0.044  }_{-0.041  }$
&
	&\phn     115$^{+18   }_{-15   }$
	&    13.7$^{+2.5 }_{-2.4 }$
	&   2380$^{+210    }_{-180    }$
	&$     55$
&
	&\phn      28$^{+15   }_{-11   }$
	&\phn     3.3$^{+1.9 }_{-1.8 }$
\\
\enddata
\tablecomments{Same as Table~\ref{tb:stronglinesA} for different
lines.}
\tablenotetext{a}{Flux and EW are the sum of the \civ\ doublet.
FWHM of \civ\ is for a single component of the doublet}
\tablenotetext{b}{\siiii\ is assumed to have the same 
FWHM and \dv\ as \ciii\ (Table~\ref{tb:fitpar}).
}
\end{deluxetable}

%%%%%%%%%%%%%%%%%%%	strong C
% \addtocounter{table}{-1}
% \addtocounter{strongline}{1}
\begin{deluxetable}{clllrp{0pt}@{}lllrrp{0pt}@{}lllr}
\tabletypesize{\scriptsize}
\rotate
\tablecolumns{16}
\tablewidth{0pc}
\tablecaption{Parameters of Strong Emission Lines 
\mgii, \hb, and \oiii
\label{tb:stronglinesC}}
\tablehead{
\colhead{} & 
\multicolumn{4}{c}{\mgii\tablenotemark{a}} & \colhead{} &
\multicolumn{5}{c}{\hb} & \colhead{} &
\multicolumn{4}{c}{\oiiil}
\\ \cline{2-5} \cline{7-11} \cline{13-16}
\colhead{Object} &
\colhead{Flux}&\colhead{EW}&\colhead{FWHM}&\colhead{\dv}&
\colhead{} &
\colhead{Flux}&\colhead{EW}&\colhead{FWHM}&\colhead{\dv}&\colhead{Asymm} &
\colhead{} &
\colhead{Flux}&\colhead{EW}&\colhead{FWHM}&\colhead{\dv}
}
\startdata
4PG0947	&\phn     122$^{+19   }_{-14   }$
	&    36.9$^{+6.7 }_{-6.4 }$
	&   4900$^{+185    }_{-175    }$
	&$   -300$
&
	&\phn     122$^{+7    }_{-6    }$
	&   121.1$^{+8.2 }_{-8.1 }$
	&   4540$^{+40     }_{-30     }$
	&$   -105$
	&$ -0.035^{+0.003  }_{-0.002  }$
&
	&\phn    10.9$^{+0.3  }_{-0.3  }$
	&    11.3$^{+0.5 }_{-0.4 }$
	&\phn    603$^{+0      }_{-10     }$
	&$      5$
\\
PG0953	&\phn     168$^{+23   }_{-18   }$
	&    28.4$^{+4.4 }_{-4.3 }$
	&   2615$^{+70     }_{-65     }$
	&$    -85$
&
	&\phn     181$^{+12   }_{-10   }$
	&\phn    95.9$^{+7.6 }_{-7.4 }$
	&   3205$^{+20     }_{-20     }$
	&$    240$
	&$ -0.020^{+0.001  }_{-0.001  }$
&
	&\phn    23.2$^{+0.7  }_{-0.6  }$
	&    13.0$^{+0.6 }_{-0.5 }$
	&\phn    721$^{+0      }_{-10     }$
	&$    -85$
\\
PG1001	&\phn \phn      62$^{+19   }_{-13   }$
	&    34.4$^{+11.8}_{-11.0}$
	&   2280$^{+90     }_{-80     }$
	&$   -155$
&
	&\phn \phn      59$^{+3    }_{-2    }$
	&\phn    88.3$^{+5.5 }_{-5.4 }$
	&   2405$^{+25     }_{-25     }$
	&$    -50$
	&$  0.059^{+0.004  }_{-0.002  }$
&
	&\phn \phn     7.5$^{+0.7  }_{-0.6  }$
	&    12.1$^{+1.3 }_{-1.2 }$
	&   1125$^{+30     }_{-30     }$
	&$    -55$
\\
PG1114	&\phn     188$^{+46   }_{-34   }$
	&    40.4$^{+11.5}_{-10.8}$
	&   4255$^{+215    }_{-205    }$
	&$   -125$
&
	&\phn     192$^{+10   }_{-8    }$
	&\phn    91.8$^{+5.7 }_{-5.6 }$
	&   4825$^{+45     }_{-40     }$
	&$     40$
	&$ -0.135^{+0.011  }_{-0.011  }$
&
	&\phn    31.3$^{+0.7  }_{-0.7  }$
	&    15.8$^{+0.5 }_{-0.5 }$
	&\phn    706$^{+10     }_{-0      }$
	&$    -95$
\\
PG1115	&\phn     110$^{+38   }_{-23   }$
	&    28.2$^{+11.1}_{-10.3}$
	&   2750$^{+140    }_{-130    }$
	&$    290$
&
	&\phn \phn      85$^{+5    }_{-4    }$
	&\phn    74.5$^{+5.9 }_{-5.7 }$
	&   1840$^{+15     }_{-15     }$
	&$    245$
	&$ -0.027^{+0.002  }_{-0.001  }$
&
	&\phn \phn     8.7$^{+0.8  }_{-0.6  }$
	&\phn     8.0$^{+0.9 }_{-0.8 }$
	&\phn    829$^{+20     }_{-10     }$
	&$     25$
\\
\enddata
\tablecomments{Same as Table~\ref{tb:stronglinesA} for different
lines.}
\tablenotetext{a}{Flux and EW are the sum of the \mgii\ doublet.
FWHM of \mgii\ is for a single component of the doublet
}
\end{deluxetable}

%\clearpage
%%%%%%%%%%%%%%%%%%%  Fe, sbb			strong D
% \addtocounter{table}{-1}
% \addtocounter{strongline}{1}
\begin{deluxetable}{cllp{0pt}@{}llp{0pt}@{}rrp{0pt}@{}rr}
\tabletypesize{\scriptsize}
\tablecolumns{16}
\tablewidth{0pc}
\tablecaption{Parameters of Strong Emission Lines \lya, \nv, and}
\tablecaption{Parameters of
UV and Optical \fe\ Blends, and the Small Blue Bump
\label{tb:stronglinesD}}
\tablehead{
\colhead{} & \colhead{} & %\colhead{} & \colhead{} &
\colhead{} & \colhead{} &
\colhead{} &
\colhead{} &
\colhead{} &
\multicolumn{5}{c}{Small Blue Bump\tablenotemark{c}}
\\ \cline{8-12}
\colhead{} & 
\multicolumn{2}{c}{Optical \feii\tablenotemark{a}} &
\colhead{} &
\multicolumn{2}{c}{UV Fe\tablenotemark{b}} & 
\colhead{} &
\multicolumn{2}{c}{local cont.} & 
\colhead{} &
\multicolumn{2}{c}{global cont.}
\\ \cline{2-3} \cline{5-6} \cline{8-9} \cline{11-12}
\colhead{Object} &
\colhead{Flux}&\colhead{EW}& %\colhead{FWHM}&
\colhead{} &
\colhead{Flux}&\colhead{EW}& %\colhead{FWHM}&
\colhead{} &
\colhead{Flux}&\colhead{EW}& %\colhead{$f_{3646}$\tablenotemark{c}}&
\colhead{} &
\colhead{Flux}&\colhead{EW} %&\colhead{$f_{3646}$\tablenotemark{c}}
}
\startdata
PG0947	&\phn     113$^{+8    }_{-8    }$
	&   111.8$^{+8.9 }_{-8.7 }$
&
	&\phn     416$^{+71   }_{-71   }$
	&   125.8$^{+25.1}_{-23.9}$
&
	&     978&   387&
	&    1300&   544
\\
PG0953	&\phn     113$^{+15   }_{-15   }$
	&\phn    60.0$^{+8.9 }_{-8.7 }$
&
	&\phn     901$^{+91   }_{-91   }$
	&   152.1$^{+18.3}_{-17.7}$
&
	&    1714&   362&
	&    3402&   875
\\
PG1001	&\phn     136$^{+9    }_{-9    }$
	&   203.4$^{+27.6}_{-24.4}$
&
	&\phn     196$^{+53   }_{-53   }$
	&   108.2$^{+34.3}_{-32.0}$
&
	&     604&   433&
	&     946&   777
\\
PG1114	&\phn     102$^{+15   }_{-15   }$
	&\phn    48.8$^{+7.8 }_{-7.7 }$
&
	&\phn     768$^{+141  }_{-141  }$
	&   164.6$^{+36.7}_{-34.5}$
&
	&    1825&   481&
	&    3669&  1295
\\
PG1115	&\phn     170$^{+14   }_{-14   }$
	&   149.0$^{+19.1}_{-17.7}$
&
	&\phn     339$^{+134  }_{-134  }$
	&\phn    86.7$^{+39.0}_{-36.1}$
&
	&    1029&   358&
	&    1178&   413
\\
PG1116	&\phn     464$^{+17   }_{-17   }$
	&   171.9$^{+10.5}_{-10.1}$
&
	&    1892$^{+196  }_{-196  }$
	&   191.0$^{+24.2}_{-23.2}$
&
	&    3677&   490&
	&    5315&   782
\\
PG1202	&\phn \phn      45$^{+9    }_{-9    }$
	&\phn    59.3$^{+12.0}_{-11.8}$
&
	&\phn     209$^{+28   }_{-28   }$
	&   113.1$^{+17.2}_{-16.6}$
&
	&     938&   711&
	&    1254&  1103
\\
PG1216	&\phn \phn      57$^{+7    }_{-7    }$
	&\phn    42.1$^{+6.2 }_{-6.0 }$
&
	&\phn     432$^{+59   }_{-59   }$
	&   175.6$^{+30.8}_{-28.9}$
&
	&     260&    99&
	&     825&   358
\\
\enddata
\tablecomments{Same as Table~\ref{tb:stronglinesA} for different
features.}
\tablenotetext{a}{Flux is from the template fitting between 4478--5640\AA.
EW is calculated using the local continuum at 4861\AA.}
\tablenotetext{b}{Flux is from the template fitting between 2230--3016\AA.
EW is calculated using the local continuum at 2799\AA.}
\tablenotetext{c}{See \S\ref{sec:measure:sbb} for descriptions.}

\end{deluxetable}

%%%%%%%%%%%%%%%%%%% HeI, Ha		strong E
% \addtocounter{table}{-1}
% \addtocounter{strongline}{1}
\begin{deluxetable}{clllrp{0pt}@{}lllrr}
\tabletypesize{\scriptsize}
\tablecolumns{11}
\tablewidth{0pc}
\tablecaption{Parameters of Strong Emission Lines 
\nai+\hei\ and \ha
\label{tb:stronglinesE}}
\tablehead{
\colhead{} & 
\multicolumn{4}{c}{\nail\ + \heil} & \colhead{} &
\multicolumn{5}{c}{\ha} 
\\ \cline{2-5} \cline{7-11} %\cline{13-15} \cline{17-19}
\colhead{Object} &
\colhead{Flux}&\colhead{EW}&\colhead{FWHM}&\colhead{\dv\tablenotemark{a}}&
\colhead{} &
\colhead{Flux}&\colhead{EW}&\colhead{FWHM}&\colhead{\dv}&\colhead{Asymm} 
}
\startdata
PG0947	&\phn    19.7$^{+6.4  }_{-4.8  }$
	&    23.8$^{+8.1 }_{-7.9 }$
	&   4655$^{+385    }_{-350    }$
	&$    -10$
&
	&\phn   338.0$^{+6.6  }_{-5.8  }$
	&   422.0$^{+19.2}_{-18.3}$
	&   3735$^{+20     }_{-15     }$
	&$    -80$
	&$ -0.086^{+0.002  }_{-0.002  }$
\\
PG0953	&\phn    18.8$^{+3.4  }_{-2.5  }$
	&    13.9$^{+2.6 }_{-2.6 }$
	&   3190$^{+80     }_{-85     }$
	&$    280$
&
	&\phn   477.3$^{+3.0  }_{-2.7  }$
	&   422.7$^{+6.5 }_{-6.3 }$
	&   2700$^{+10     }_{-0      }$
	&$     95$
	&$  0.001^{+0.001  }_{-0.000  }$
\\
PG1001	&\phn \phn     8.4$^{+1.5  }_{-1.2  }$
	&    17.0$^{+3.1 }_{-3.1 }$
	&   3010$^{+160    }_{-145    }$
	&$    300$
&
	&\phn   199.7$^{+1.7  }_{-1.5  }$
	&   457.7$^{+14.7}_{-14.0}$
	&   1730$^{+0      }_{-0      }$
	&$     80$
	&$  0.075^{+0.000  }_{-0.003  }$
\\
PG1114	&\phn    36.7$^{+8.9  }_{-7.1  }$
	&    22.5$^{+5.8 }_{-5.7 }$
	&   5310$^{+565    }_{-480    }$
	&$    355$
&
	&\phn   720.6$^{+8.1  }_{-7.2  }$
	&   485.6$^{+8.8 }_{-8.7 }$
	&   3955$^{+10     }_{-10     }$
	&$    150$
	&$ -0.024^{+0.002  }_{-0.002  }$
\\
PG1115	&\phn    17.2$^{+2.1  }_{-1.7  }$
	&    17.5$^{+2.3 }_{-2.3 }$
	&   2300$^{+40     }_{-50     }$
	&$    505$
&
	&\phn   317.8$^{+4.0  }_{-3.5  }$
	&   355.4$^{+8.5 }_{-8.3 }$
	&   1690$^{+0      }_{-5      }$
	&$    260$
	&$ -0.006^{+0.000  }_{-0.001  }$
\\
PG1116	&\phn    47.7$^{+4.2  }_{-3.5  }$
	&    23.6$^{+2.2 }_{-2.2 }$
	&   3535$^{+60     }_{-60     }$
	&$    180$
&
	&  1153.2$^{+8.6  }_{-7.7  }$
	&   646.6$^{+12.2}_{-12.0}$
	&   2720$^{+0      }_{-0      }$
	&$    235$
	&$  0.019^{+0.001  }_{-0.001  }$
\\
PG1202	&\phn    18.1$^{+2.8  }_{-2.3  }$
	&    28.6$^{+5.0 }_{-4.8 }$
	&   3795$^{+690    }_{-785    }$
	&$     35$
&
	&\phn   334.8$^{+4.6  }_{-4.1  }$
	&   559.1$^{+17.4}_{-16.8}$
	&   4015$^{+10     }_{-5      }$
	&$   -300$
	&$ -0.035^{+0.000  }_{-0.002  }$
\\
\enddata
\tablecomments{Same as Table~\ref{tb:stronglinesA} for different
lines.}
\tablenotetext{a}{Calculated using \heil\ wavelength.}
\end{deluxetable}

%%%%%%%%%%%%%%%%%%%%%%%%%%%%%%%%%%%%%%%%%%%%%%%%%%%%% Weak Line Measurements

\renewcommand{\thetable}{\arabic{table}}

%%%%%%%%%%%%%%%%%%%%%%%%%%%%%%%%%%%%%%%%%%%%%%%%%%%%%%%%%% Mbh
\begin{deluxetable}{lrcccccc}
%\tabletypesize{\small}
\tabletypesize{\scriptsize}
\tablecolumns{8}
\tablewidth{0pc}
\tablecaption{Luminosity, Black Hole Mass, and Eddington Ratio
\label{tb:mbh}
}
\tablehead{
\colhead{}&
\colhead{}&
\colhead{\llciv\tablenotemark{a}}&
\colhead{FWHM(\hb)}&
\colhead{}&
\colhead{\lbol\tablenotemark{a}}&
\colhead{\mbh}&
\colhead{}\\
\colhead{Object}&
\colhead{$z$}&
\colhead{[log(erg s$^{-1}$)]}&
\colhead{(\kms)}&
\colhead{$f_\lambda(5100{\rm \AA})$\tablenotemark{b}}&
\colhead{[log(erg s$^{-1}$)]}&
\colhead{($10^8 M_\odot$)}&
\colhead{\lledd}
}
\startdata
PG0947+396& 0.2056&  45.16&   4540&   0.94&  45.66&   2.34&   0.16 \\
PG0953+414& 0.2341&  45.54&   3205&   1.71&  46.05&   2.19&   0.41 \\
PG1001+054& 0.1603&  44.54&   2405&   0.60&  45.22&   0.32&   0.42 \\
PG1114+445& 0.1440&  44.66&   4825&   1.93&  45.63&   2.52&   0.14 \\
PG1115+407& 0.1541&  44.98&   1840&   1.06&  45.43&   0.27&   0.81 \\
PG1116+215& 0.1759&  45.55&   2975&   2.42&  45.92&   1.53&   0.44 \\
PG1202+281& 0.1651&  44.37&   4950&   0.70&  45.32&   1.61&   0.10 \\
PG1216+069& 0.3319&  45.45&   5950&   1.23&  46.26&  10.56\phn&   0.14 \\
\enddata
%\tablecomments{}
\tablenotetext{a}{ Assuming zero cosmological constant, \Hoeq,
and $q_0=0.5$, same as the cosmology used in \citet{Kasp00}.}
\tablenotetext{b}{Observed-frame flux density at 5100\,\AA\ in the
units of $10^{-15}$\fluxl, from fitted local continuum in \hb\ region.}
\end{deluxetable}

%%%%%%%%%%%%%%%%%%%%%%%%%%%%%%%%%%%%%%%%%%%%%%%%%%%%%%%%%%%%
%%%%%%%%%%%
%%%%%%%%%%% Correlation tables

%%%%%%%%%%%%%%%%%%%%%%%%%%%%%%
%%%%%%%%%%%%%%%%%%%%%%%%%%%%%%	
\begin{deluxetable}{lrrr}
%\tabletypesize{\scriptsize}
\tablecolumns{4}
\tablewidth{0pc}
\tablecaption{Statistics of Emission-Line Shifts (\kms)
\label{tb:shift}
}
\tablehead{
\colhead{Line}&\colhead{Mean}&\colhead{$\sigma$}&\colhead{Median}
}
\startdata
\lya	&	 -90 &      250 &      -94  \\
\civ	&	-445&       513 &     -458  \\
\civ\tablenotemark{a}	&	-368&       373 &     -449  \\
\ciii	&	 -16 &      225 &        2  \\
\mgii	&	 -71 &      192 &      -69  \\
\hb	&	  91  &     163 &      113  \\
\ha	&	 147 &      161 &      170  \\
%# nrows            mean        stddev        median           min
%# max
%   22      -90.35909091       249.883        -93.65        -549.2
%   22      -444.6086364       513.494      -457.865      -2063.33
%civ2   21      -367.5266667       373.643       -448.91        -870.8
%   21      -15.79047619       224.697           1.6        -781.1
%   22      -70.70454545       191.809        -68.65        -533.2
%   22       91.21363636       163.019        113.05         -253.
%   22       146.9304545       161.032        169.62       -299.55

%       &      1  &      2  &      3  &      4  &      5  \\
%
%
\enddata
\tablecomments{For the whole sample.  $\sigma$ is the standard
deviation of the mean.}
\tablenotetext{a}{Excluding one extreme blueshift of $-$2065 \kms\
(PG1543+489).}
\end{deluxetable}
%%%%%%%%%%%

%%%%%%%%%%%%%%%%%%%%%%%%%%%%%%	Shift 
%%%%%%%%%%%%%%%%%%%%%%%%%%%%%%	Shift 
%%%%%%%%%%%%%%%%%%%%%%%%%%%%%%	
\begin{deluxetable}{lrrrrrr}
\tabletypesize{\scriptsize}
%\tabletypesize{\tiny}
\tablecolumns{12}
\tablewidth{0pc}
\tablecaption{Pearson Correlation Coefficients for Emission-Line
Velocity Shift 
\label{tb:corrshift}
}
\tablehead{
\colhead{} &
\multicolumn{6}{c}{Velocity Shift $\Delta v$} \\ \cline{2-7}
\colhead{}&\colhead{\lya}&\colhead{\civ}&\colhead{\ciii\tablenotemark{a}}&\colhead{\mgii}&\colhead{\hb}&\colhead{\ha}
}
\startdata
 \lyamsh&        &    0.81&    0.45&    0.06&    0.24&    0.10\\
\civmshi&    0.81&        &    0.39& $-$0.11&    0.11&    0.03\\
\ciiishi\tablenotemark{a}&    0.45&    0.39&        &    0.52&    0.47&    0.53\\
\mgiishi&    0.06& $-$0.11&    0.52&        &    0.56&    0.36\\
\hbmshif&    0.24&    0.11&    0.47&    0.56&        &    0.62\\
\hamshif&    0.10&    0.03&    0.53&    0.36&    0.62&           \\
\enddata
\tablecomments{22 objects are used unless noted.  
A significance level of $>99\%$ for a correlation 
requires a correlation coefficient $>0.53$.}
\tablenotetext{a}{\ciii\ is not available for PG~1543+489, 21 objects are used.}
\end{deluxetable}
%%%%%%%%%%%%%%%%%%%%%%%%%%%%%%	Shift 
%%%%%%%%%%%%%%%%%%%%%%%%%%%%%%	Shift 

\clearpage

%%%%%%%%%%%%%%%%%%%%%%%%%%%%%%	 Asymmetry
%%%%%%%%%%%%%%%%%%%%%%%%%%%%%%	 Asymmetry
%%%%%%%%%%%%%%%%%%%%%%%%%%%%%%	
\begin{deluxetable}{lrrrr}
\tabletypesize{\scriptsize}
%\tabletypesize{\tiny}
\tablecolumns{12}
\tablewidth{0pc}
\tablecaption{Pearson Correlation Coefficients for Emission-Line
Asymmetry
\label{tb:corrasymm}
}
\tablehead{
\colhead{} &
%\multicolumn{6}{c}{Shift $\Delta v$} &
%\colhead{} &
\multicolumn{4}{c}{Asymmetry} \\ \cline{2-5} 
\colhead{}&\colhead{\lya}&\colhead{\civ}&\colhead{\hb}&\colhead{\ha} 
}
\startdata
\lyaasym&         &    0.53& $-$0.30&	 0.16\\
\civasym&     0.53&        & $-$0.25&	 0.35\\
\hbasymm&  $-$0.30& $-$0.25&        &    0.43 \\
\haasymm&     0.16&    0.35&    0.43&          \\
\enddata
\tablecomments{22 objects are used.
A significance level of $>99\%$ for a correlation
requires a correlation coefficient $>0.53$.}
\end{deluxetable}


\begin{thebibliography}{}

%\input{ref.tex}

\bibitem[Bahcall et al.(1997)Bahcall, Korhakos \&
Saxe]{Bahc97}Bahcall, J.~N., Kirhakos, S., \& Saxe, D.~H. 1997, \apj,
479, 642

\bibitem[Baldwin(1977)]{Bald77}Baldwin, J.~A.\ 1977, ApJ, 214, 679

\bibitem[Baldwin et al.(1989)]{Bald89} Baldwin, J.~A., 
Wampler, E.~J., \& Gaskell, C.~M.\ 1989, \apj, 338, 630 

\bibitem[Baldwin et al.(1996)]{Bald96}Baldwin, J.~A. et al. 1996,
\apj, 461, 664

\bibitem[Baskin \& Laor(2004)]{BasLao04} Baskin, A. \& Laor, A. 2004,
\mnras, 350, L31 

\bibitem[Baskin \& Laor(2005)]{BasLao05} Baskin, A. \& Laor, A. 2005,
\mnras, 356, 1029

\bibitem[Bohlin(1996)]{Bohl96}Bohlin, R.~C.\ 1996, AJ, 111, 1743

\bibitem[Bohlin(2000)]{Bohl00}Bohlin, R.~C.\ 2000, AJ, 120, 437

\bibitem[Boroson \& Green(1992)]{BG92}Boroson, T.~A, \& Green,
R.~F.\ 1992, ApJS, 80, 109 (BG92)

\bibitem[Boroson(2002)]{Boro02} Boroson, T.~A.\ 2002, \apj, 565, 78
(B02)

\bibitem[Boroson(2005)]{Boro05} Boroson, T.~A.\ 2005, \aj, 130, 381

%\bibitem[Brinkmann, Yuan, \& Siebert(1997)]{Brin97} Brinkmann, W.,
%Yuan, W., \& Siebert, J. 1997, \aap, 319, 413

%\bibitem[Brotherton et al.(1994a)]{Brot94a}Brotherton, M.~S., Wills,
%B.~J., Francis, P.~J., \& 
% Steidel, C.~S.\ 1994a, \apj, 430, 495
%The intermediate line region of QSOs

%\bibitem[Brotherton et al.(1994b)]{Brot94b}
% Brotherton, M.~S., Wills, B.~J., Steidel, C.~C., \& Sargent,
% W.~L.~W.\ 1994b, \apj, 423, 131
%broad emission-line profiles. 2: The C IV wavelength 1549, C III)
%wavelength 1909, and MG II wavelength 2798 lines

\bibitem[Cardelli et al.(1989)Cardelli, Clayton, \& Mathis]{Card89}
Cardelli, J.~A., Clayton, G.~C., \& Mathis, J.~S., 1989, \apj, 345,
245

\bibitem[Corbin(1990)]{Corb90} Corbin, M.~R. 1990, \apj, 357, 346

\bibitem[Corbin(1991)]{Corb91} Corbin, M.~R.\ 1991, \apjl, 371, L51 

\bibitem[Corbin \& Boroson(1996)]{CorBor96} Corbin, M.~R. \& Boroson,
T.~A.  1996, \apj, 107, 69

\bibitem[Croom et al.(2002)]{Croo02}Croom, S.~M. et al. 2002, \mnras,
337, 275

\bibitem[De Robertis(1985)]{DeRo85}De Robertis, M. M. 1985, \apj,
289, 67

\bibitem[Dietrich et al.(2002)]{Diet02} Dietrich, M.,
Hamann, F., Shields, J.~C., Constantin, A., Vestergaard, M., Chaffee,
F., Foltz, C.~B., \& Junkkarinen, V.~T.\ 2002, \apj, 581, 912 

\bibitem[Dunlop et al.(2003)]{Dunl03} Dunlop, J.~S., McLure, R.~J.,
Kukula, M.~J., Baum, S.~A., O'Dea, C.~P., \& Hughes, D.~H. 2003,
\mnras, 340, 1095

\bibitem[Elvis et al.(1994)]{Elvi94} Elvis, M., Wilkes, B.~J.,
McDowell, J.~C., Green, R.~F., Bechtold, J., Willner, S.~P., Oey,
M.~S., Polomski, E., \& Cutri, R. 1994, \apjs, 95, 1

\bibitem[Espey et al.(1989)]{Espe89} Espey, B.~R., Carswell, R.~F.,
Bailey, J.~A., Smith, M.~G., \& Ward, M.~J.\ 1989, \apj, 342, 666 

\bibitem[Espey \& Andreadis(1999)]{EspAnd99} Espey, B., \&
Andreadis, S.\ 1999, ASP Conf.~Ser.~162: Quasars and Cosmology, 162,
351 

%\bibitem[Elvis(2000)]{Elvi00}
%Elvis, M. 2000, \apj, 545, 63

\bibitem[Fabian et al.(2006)]{Fab06} Fabian, A.~C., Celotti, 
A., \& Erlund, M.~C.\ 2006, \mnras, L98 

\bibitem[Ferland et al.(1979)]{Fer79} Ferland, G.~J., 
Shields, G.~A., \& Netzer, H.\ 1979, \apj, 232, 382 

\bibitem[Francis et al.(1992)]{Fran92} Francis, P.~J., Hewett, P.~C.,
Foltz, C.~B., \& Chaffee, F.~H.\ 1992, ApJ, 398, 476  

%\bibitem[Francis(1993)]{Fran93} Francis, P.~J. 1993, \apj, 407, 519

\bibitem[Francis \& Wills (1999)]{FraWil99} Francis, P.~J. \&
Wills, B. J. 1999, in ASP Conf. Series 162, Quasars and Cosmology, ed.
G.~J. Ferland, \& J.~A. Baldwin (San Francisco: ASP), 363

%\bibitem[Gallagher et al.(2005)]{Gal05} Gallagher, S.~C., 
%Richards, G.~T., Hall, P.~B., Brandt, W.~N., Schneider, D.~P., \& Vanden 
%Berk, D.~E.\ 2005, \aj, 129, 567 


\bibitem[Gaskell (1982)]{Gask82} Gaskell, C.~M.\ 1982,
\apj, 263, 79 

\bibitem[Green et al.(2001)]{Gree01} Green, P.~J.,
Forster, K., \& Kuraszkiewicz, J.\ 2001, \apj, 556, 727 

\bibitem[Greene \& Ho(2005)]{GreHo05} Greene, J.~E., \&
Ho, L.~C.\ 2005, \apj, 630, 122

\bibitem[Jester et al.(2005)]{Jes05} Jester, S., et al.\ 
2005, \aj, 130, 873 

\bibitem[Kaspi et al.(2000)]{Kasp00} Kaspi, S., Smith, P.~S., Netzer,
H., Maoz, D., Jannuzi, B.~T., \& Giveon, U.\ 2000, \apj, 533, 631

\bibitem[Keyes et al.(1995)]{Key95} Keyes, C.~D., Koratkar, A.~P., Dahlem, M., Hayes, J., Christensen, J., \& Martin, S. 1995, {http://www.stsci.edu/hst/HST\_overview/documents}

\bibitem[Kinney et al.(1990)Kinney, Rivolo \& Koratkar]{Kinn90}
Kinney, A.~L., Rivolo, A.~R., \&
Koratkar, A.~P.\ 1990, \apj, 357, 338

\bibitem[Korista et al.(1995)]{Kori95}Korista, K.~T. et al.
1995, \apjs, 97, 285

\bibitem[Kriss(1994)]{Kris94}Kriss, G.~A. 1994, in ASP Conf. Series
61, Third Conference on Astrophysics Data Analysis and Software
Systems III, ed. D.~R. Crabtree, R.~J. Hanisch \& J. Barnes (ASP:San
Francisco), 437

\bibitem[Kuraszkiewicz et al.(2002)]{Kura02}
Kuraszkiewicz, J.~K., Green, P.~J., Forster, K., Aldcroft, T.~L.,
Evans, I.~N., \& Koratkar, A.\ 2002, \apjs, 143, 257 

\bibitem[Laor et al.(1994a)]{Laor94a} Laor, A., Bahcall,
J.~N., Jannuzi, B.~T., SchneiDER, d.~p., Green, R.~F., \& Hartig,
G.~F.\ 1994, \apj, 420, 110 
%The ultraviolet emission properties of 5 low-redshift active
%galactic nuclei at high signal-to-noise ratio and spectral resolution

\bibitem[Laor et al.(1994b)]{L94}Laor, A., Fiore, F., Elvis, M.,
Wilkes, B.~J., \& McDowell, J.~C.\ 1994, \apj, 435, 611 (L94)
%The soft x-ray properties of a complete sample of optically selected
%quasars. 1: First results

\bibitem[Laor et al.(1995)]{Laor95}Laor, A., Bahcall, J.~N., Jannuzi,
B.~T., Schneider, D.~P., \& Green, R.~F.\ 1995, \apjs, 99, 1
%The Ultraviolet Emission Properties of 13 Quasars

\bibitem[Laor et al.(1997a)]{L97} Laor, A., Fiore, F., Elvis, M.,
Wilkes, B.~J., \& McDowell, J.~C.\ 1997a, \apj, 477, 93 (L97)

\bibitem[Laor et al.(1997b)]{Laor97b} Laor, A., Jannuzi, B.~T.,
Green, R.~F., \& Boroson, T.~A. 1997b, \apj, 489, 656
%The Ultraviolet Properties of the Narrow-Line Quasar I ZW 1


\bibitem[Leighly \& Moore(2004)]{LeiMoo04} Leighly K. M. \& Moore J.
R.  2004, \apj, 611, 107

\bibitem[Leighly(2004)]{Leig04} Leighly, K.~M.\ 2004,
\apj, 611, 125 

\bibitem[Malkan \& Moore(1986)]{MaMo86} Malkan, M.~A., \& 
Moore, R.~L.\ 1986, \apj, 300, 216 

\bibitem[Marconi \& Hunt(2003)]{MarHun03} Marconi, A., \& Hunt, 
L.~K.\ 2003, \apjl, 589, L21 

\bibitem[Marziani et al.(1996)]{Marz96} Marziani, P., 
Sulentic, J.~W., Dultzin-Hacyan, D., Calvani, M., \& Moles, M.\ 1996, 
\apjs, 104, 37 

\bibitem[Sulentic et al.(2000)]{Sule00} Sulentic, J.~W.,             
Marziani, P., \& Dultzin-Hacyan, D.\ 2000, \araa, 38, 521
 
\bibitem[McLeod \& Rieke(1994a)]{McLRie94a} McLeod, K.~K., \&
Rieke, G.~H. 1994a, \apj, 420, 58

\bibitem[McLeod \& Rieke(1994b)]{McLRie94b} McLeod, K.~K., \&
Rieke, G.~H. 1994b, \apj, 431, 137

\bibitem[McLure \& Jarvis(2002)]{McLJar02} McLure, R.~J.,
\& Jarvis, M.~J.\ 2002, \mnras, 337, 109 

%\bibitem[Nagao et al.(2006)]{Nag06} Nagao, T., Marconi, A., 
%\& Maiolino, R.\ 2006, \aap, 447, 157 

%\bibitem[Natali et al.(1998)]{Nata98}Natali, F., Giallongo, E., 
%Cristiani, S., \& Franca, F.~L.\ 1998, \aj, 115, 397

\bibitem[O'Brien et al.(1988)]{OBri88} O'Brien, P.~T., Wilson, 
R., \& Gondhalekar, P.~M.\ 1988, \mnras, 233, 801 

\bibitem[Peterson et al.(1991)]{Pete91}Peterson, B.~M. et al.
1991, \apj, 368, 119

\bibitem[Peterson(1993)]{Pete93} Peterson, B.~M.\ 1993,
\pasp, 105, 247 

\bibitem[Peterson et al.(2004)]{Pete04} Peterson, B.~M.,
et al.\ 2004, \apj, 613, 682 

%\bibitem[Pfefferkorn, Boller, \& Rafanelli(2001)]{Pfef01}
%Pfefferkorn, F., Boller, Th., \& Rafanelli, P. 2001, \aap, 368, 797

%\bibitem[Pounds et al.(1995)]{PDO95} Pounds, K.~A., Done, C., 
%\& Osborne, J.~P.\ 1995, \mnras, 277, L5 

\bibitem[Richards et al.(2002)]{Rich02} Richards, G.~T., 
Vanden Berk, D.~E., Reichard, T.~A., Hall, P.~B., Schneider, D.~P., 
SubbaRao, M., Thakar, A.~R., \& York, D.~G.\ 2002, \aj, 124, 1 

\bibitem[Richards et al.(2006a)]{Richa06} Richards, G.~T.,
et al.\ 2006, \apjs, 166, 470 

\bibitem[Richards(2006b)]{Ric06} Richards, G.~T.\ 2006, ArXiv 
Astrophysics e-prints, arXiv:astro-ph/0603827

\bibitem[Rosa et al.(2001)]{Rosa01} Rosa, M., Alexov, A., Bristow, P.
\& Kerber, F. 2001, ST-ECF Newletter (July, 2001), 29, 9,
(\url{http://www.stecf.org/newsletter/stecf-nl-29.pdf})

\bibitem[Schlegel, Finkbeiner, \& Davis(1998)]{Schl98}
Schlegel, D. J., Finkbeiner, D. P., \& Davis, M. 1998, \apj, 500, 525

\bibitem[Schmidt \& Green(1983)]{SchGre83}Schmidt, M., \& Green,
R.~F.\ 1983, \apj, 269, 352

\bibitem[Shang et al.(2003)]{Shan03} Shang, Z., Wills, B. J.,
Robinson, E. L., Wills, D., Laor, A., Xie, B., \& Yuan, J. 2003, \apj,
586, 52

\bibitem[Shang et al.(2005)]{Shan05} Shang, Z, Brotherton, M. S.,
Green, R. F., Kriss, G. A., Scott, J., Quijano, J. K., Blaes, O.,
Hubeny, I., Hutchings, J., Kaiser, M. E., Koratkar, A., Oegerle, W.,
Zheng, W.  \ 2005, \apj, 619, 41 

%\bibitem[Shemmer et al.(2004)]{Shem04} Shemmer, O., Netzer, 
%H., Maiolino, R., Oliva, E., Croom, S., Corbett, E., \& di Fabrizio, L.\ 
%2004, \apj, 614, 547 

%\bibitem[Strateva et al.(2005)]{Strat05} Strateva, I.~V., 
%Brandt, W.~N., Schneider, D.~P., Vanden Berk, D.~G., \& Vignali, C.\ 2005, 
%\aj, 130, 387 

%\bibitem[Steffen et al.(2006)]{Stef06} Steffen, A.~T., 
%Strateva, I., Brandt, W.~N., Alexander, D.~M., Koekemoer, A.~M., Lehmer, 
%B.~D., Schneider, D.~P., \& Vignali, C.\ 2006, \aj, 131, 2826 

\bibitem[Sitko et al.(1993)]{Sitko93} Sitko, M.~L., Sitko, 
A.~K., Siemiginowska, A., \& Szczerba, R.\ 1993, \apj, 409, 139 

\bibitem[Sulentic et al.(1995)]{Sule95} Sulentic, J.~W.,
Marziani, P., Dultzin-Hacyan, D., Calvani, M., \& Moles, M.\ 1995,
\apjl, 445, L85 

\bibitem[Sulentic \& Marziani(1999)]{Sule99} Sulentic, J.~W., \&
Marziani, P.\ 1999, \apjl, 518, L9

\bibitem[Sun \& Malkan(1989)]{SuMa89} Sun, W.-H., \& Malkan, 
M.~A.\ 1989, \apj, 346, 68

\bibitem[Tremaine et al. (2002)]{Trem02} Tremaine, S., et al.
2002, \apj, 574, 740

\bibitem[Tytler \& Fan(1992)]{TytFan92} Tytler, D. \& Fan, Xiao-Ming\
1992, \apjs, 79, 1

\bibitem[Vanden Berk et al.(2001)]{Vand01} Vanden Berk, D. et al.\
2001, \aj, 122, 549

\bibitem[Vestergaard (2002)]{Vest02}  Vestergaard, M. 2002, \apj, 571,
733

\bibitem[Vestergaard(2004)]{Ves04} Vestergaard, M. 2004, \apj, 601, 676

\bibitem[Vestergaard \& Wilkes(2001)]{VesWil01} Vestergaard, M. \&
Wilkes, B.\ 2001, \apjs, 134, 1
%Fe template

\bibitem[Vestergaard \& Peterson(2006)]{VesPet06} Vestergaard, 
M., \& Peterson, B.~M.\ 2006, \apj, 641, 689


\bibitem[Wandel et al.(1999)]{Wand99} Wandel, A., Peterson, 
B.~M., \& Malkan, M.~A.\ 1999, \apj, 526, 579

%\bibitem[Warner et al.(2004)]{Warn04} Warner, C., Hamann, F., 
%\& Dietrich, M.\ 2004, \apj, 608, 136 

\bibitem[Wilkes(1984)]{Wilk84} Wilkes, B.~J.\ 1984, \mnras, 207, 73 

\bibitem[Wills(1991)]{Wil91} Wills, B.~J.\ 1991, Variability 
of Active Galactic Nuclei, 87 

\bibitem[Wills et al.(1985)Wills, Netzer, \& Wills]{Will85}
Wills, B.~J., Netzer, H., \& Wills, D. 1985, \apj, 288, 94

\bibitem[Wills et al.(1993)]{Will93} Wills, B.~J., Brotherton, M.~S.,
Fang, D., Steidel, C.~C., \& Sargent, W.~L.~W.\ 1993, \apj, 415, 563

\bibitem[Wills et al.(1995)]{Will95} Wills, B.~J., et
al.\ 1995, \apj, 447, 139 

\bibitem[Wills et al.(1999a)]{Will99a} Wills, B.~J., Laor, A.,
Brotherton, M.~S., Wills, D., Ferland, G.~J., \& Shang,
Zhaohui\ 1999a, ApJ, 515, L53

\bibitem[Wills et al.(1999c)]{Will99c} Wills, B.~J., Brotherton,
M.~S., Laor, A., Wills, D., Wilkes, B.~J., Ferland, G.~J., \& Shang,
Zhaohui\  1999b, in ASP Conf. Series 162, Quasars and Cosmology, ed.
G.~J. Ferland, \& J.~A. Baldwin (San Francisco: ASP), 373

\bibitem[Wills et al.(1999b)]{Will99b} Wills, B.~J., Brotherton,
M.~S., Laor, A., Wills, D., Wilkes, B.~J., \& Ferland, G.~J.\ 1999c,
in ASP Conf.~Ser.~175, Structure and Kinematics of Quasar Broad Line
Regions, ed. C. M. Gaskell, W. N. Brandt, M.  Dietrich, D.
Dultzin-Hacyan, \& M. Eracleous (San Francisco: ASP), 241

\bibitem[Wills et al.(2000)]{Wil00} Wills, B.~J., Shang, Z., 
\& Yuan, J.~M.\ 2000, New Astronomy Review, 44, 511

\bibitem[Wills et al. (2007)]{Will07} Wills, B.~J. et al. 2007, in
preparation

%\bibitem[Yee(1980)]{Yee80} Yee, H.K.C.\ 1980, \apj, 241, 894

\bibitem[Yuan et al.(2007)]{Yuan07} Yuan, Qirong et al. 2007, in
preparation

\bibitem[Zamanov et al.(2002)]{Zama02} Zamanov, R., Marziani, P.,
Sulentic, J. W., Calvani, M., Dultzin-Hacyan, D., \& BAchev, R. 2002,
\apj, 576, L9


\end{thebibliography}
\end{document}